\documentclass[USenglish,twocolumn]{article}
\usepackage{etex}
\usepackage[utf8]{inputenc}%(only for the pdftex engine)
\usepackage[big]{dgruyter_NEW}
\usepackage[round]{natbib}
\usepackage{latexsym}
\usepackage{enumerate}\usepackage{colortbl}
\usepackage{inputenc,epstopdf,subfig}
\inputencoding{utf8}
\usepackage{slashbox}
\usepackage{amsfonts,hypcap}
\usepackage[T1]{fontenc}
 \usepackage{multirow}
\usepackage[table]{xcolor} 
\usepackage{rotating} 
\usepackage{lscape}
\usepackage{longtable}
\usepackage{multicol}
\definecolor{light-gray}{gray}{0.95}

\usepackage{color}

\newcommand{\Puna}{\color{red}} %wczeœniej jako kolor czerwony
\usepackage[absolute]{textpos}
\usepackage[resetfonts]{cmap}
\usepackage{textcomp} 
\usepackage{appendix}
\makeatletter
\def\maketag@@@#1{\hbox{\m@th\normalfont\normalsize#1}}
\makeatother %zeby po zmniejszeniu czcionki we wzorze numer wzoru nie ulegl zmniejszeniu

\usepackage{enumitem,tikz }
\usetikzlibrary{arrows}
\setlist[itemize]{topsep=5pt,leftmargin=25pt}
\setlist[enumerate]{topsep=5pt, leftmargin= 25pt}
\usepackage{acro}

\setcitestyle{aysep={}} %do \citep

\immediate\write18{makeindex \jobname.nlo -s nomencl.ist -o \jobname.nls}
\usepackage{nomencl} 
\makenomenclature
\makeindex

\DeclareSymbolFont{arrows}{OMS}{cmsy}{m}{n}
\DeclareMathSymbol{\leftrightarrow}{\mathrel}{arrows}{"24}
\DeclareMathSymbol{\leftarrow}{\mathrel}{arrows}{"20}
   
\DeclareMathSymbol{\rightarrow}{\mathrel}{arrows}{"21}
   \let\to=\rightarrow
\DeclareMathSymbol{\mapstochar}{\mathrel}{arrows}{"37}
\DeclareMathSymbol{\relbardash}{\mathbin}{arrows}{"00}

\acsetup{first-style=short}
\DeclareMathSymbol{\prime}{\mathord}{symbols}{"30}
\newcommand{\LJnow}[1]{{\color{black}#1}}
\newcommand{\Egy}{{{\color{black} $^{{\blacktriangle}}$}}}
\newcommand{\SW}[1]{ {\it $\!\!${``#1''}}}
\newcommand{\apj}{ApJ}
\newcommand{\apjl}{The Astrophysical Journal, Letters~}
\newcommand{\mnras}{Monthly Notices of Royal Astronomical Society~}
\newcommand{\bain}{Bulletin of the Astronomical Institutes of the Netherlands}
\newcommand{\Mydryad}
{\href{http://dx.doi.org/10.5061/dryad.tj4qg}
{http://dx.doi.org/10.5061/dryad.tj4qg}}
\newcommand{\aste}{^{\mathrm{o}}}
\newcommand{\answer}[1]{{\\ {\bf Answer:} #1}}
\allowdisplaybreaks
\begin{document}
\begin{textblock}{165}(33,38)
%\begin{textblock}{165}(213,38)
%\includegraphics[width=0.35cm]{klodkamarg}
\end{textblock}
\begin{textblock}{165}(47,283)
\noindent \sffamily\scriptsize 
Open Access. \sffamily\scriptsize© 2018   S. Porceddu \textit{et al.},  published by De Gruyter. 
%\authorimg{by-nc-nd.pdf} 
This work is licensed under the Creative Commons Attribution-NonCommercial-NoDerivatives 4.0 License
\end{textblock}

\articletype{Research Article{\hfill}}
\title{\huge Algol as Horus  in the Cairo Calendar:  the possible means and the motives of the observations%\protect\footnotemark
}
\runningtitle{Algol as Horus  in the Cairo Calendar} 
\author*[1]{Sebastian Porceddu}
\author[1]{Lauri Jetsu}
\author[1]{Tapio Markkanen}
\author[1]{Joonas Lyytinen}
\author[1]{Perttu Kajatkari}
\author[2]{Jyri Lehtinen}
\author[3]{Jaana Toivari-Viitala}

\runningauthor{S. Porceddu \textit{et al.}}

\received{Feb 15, 2018}
%\revised{}
\accepted{May 04, 2018}

 \journalname{Open Astron.}
  \journalyear{2018}
  \journalvolume{27}
 \startpage{232}
   \DOI{\url{https://doi.org/10.1515/astro-2018-0033}}

\begin{abstract}
{An ancient Egyptian Calendar of Lucky and Unlucky Days, 
the Cairo Calendar (CC), assigns luck with the period of 2.850 days. 
Previous astronomical, astrophysical and statistical analyses of CC
support the idea that this was the period of the eclipsing binary Algol
three millennia ago. However, next to nothing
is known about who recorded Algol's period into CC and especially how.
Here, we show that the ancient Egyptian scribes had the possible
means and the motives for such astronomical observations. 
Their principles of describing celestial phenomena as 
activity of gods reveal 
why Algol received the title of Horus}
\end{abstract}
\maketitle

{ \let\thempfn\relax% Remove footnote number printing mechanism 
\footnotetext{\hspace{-1ex}{\Authfont\small \textbf{Corresponding Author: Sebastian Porceddu:}} {\Affilfont Department of Physics, 
            University of Helsinki, Finland;\newline
Email: sebastian.porceddu@helsinki.fi}}
}

{ \let\thempfn\relax% Remove footnote number printing mechanism 
\footnotetext{\hspace{-1ex}{\Authfont\small \textbf{Lauri Jetsu:}} {\Affilfont Department of Physics, University of Helsinki, Finland}}
}

{ \let\thempfn\relax% Remove footnote number printing mechanism 
\footnotetext{\hspace{-1ex}{\Authfont\small \textbf{Tapio Markkanen:}} {\Affilfont Department of Physics, University of Helsinki, Finland}}
}

{ \let\thempfn\relax% Remove footnote number printing mechanism 
\footnotetext{\hspace{-1ex}{\Authfont\small \textbf{Joonas Lyytinen:}} {\Affilfont Department of Physics, University of Helsinki, Finland}}
}

{ \let\thempfn\relax% Remove footnote number printing mechanism 
\footnotetext{\hspace{-1ex}{\Authfont\small \textbf{Perttu Kajatkari:}} {\Affilfont Department of Physics, University of Helsinki, Finland}}
}

{ \let\thempfn\relax% Remove footnote number printing mechanism 
\footnotetext{\hspace{-1ex}{\Authfont\small \textbf{Jyri Lehtinen:}} {\Affilfont Max-Planck-Institut f\"ur Sonnensystemforschung,
             G\"ottingen, Germany}}
}

{ \let\thempfn\relax% Remove footnote number printing mechanism 
\footnotetext{\hspace{-1ex}{\Authfont\small \textbf{Jaana Toivari-Viitala:}} {\Affilfont Department of World Cultures, 
            University of Helsinki, Finland}}
}

%\nomenclature{bpz}{2,2-bipyrazine}
\newcolumntype{Y}{>{\centering\arraybackslash}X}
\newcolumntype{P}{>{\centering\arraybackslash}p}

\section{Introduction}
 \renewcommand{\figurename}{Figure}
 
The ancient Egyptian texts known as the 
Calendars of Lucky and Unlucky Days, or hemerologies, 
are literary works that assign prognoses 
to each day of the Egyptian year 
\citep[][\LJnow{p117-118}]{Wel01}, 
\citep[][\LJnow{p1-2}]{Lei94}
\citep[][\LJnow{p41-45}]{Bac90}
\citep[][\LJnow{p127-147}]{Tro89}
and
\citep[][\LJnow{p156}]{Hel75}. 
These prognoses denote whether the day, or a part of the day, 
is considered  ``good''\Egy or ``bad''\Egy.
\footnote{
We use the symbol ``\Egy'' to denote the words and phrases
translated into Ancient Egyptian language in the list of
Appendix A.}
Nine such texts have been found
\citep[][\LJnow{p140-143}]{Tro89},
\citep[][\LJnow{p-2}]{Lei94}
and
\citep[][\LJnow{p328}]{Por08}. % RefOK
Here, we study the best preserved one of these nine texts, 
CC, dated to 1271-1163 B.C. 
\citep[][\LJnow{p2-5}]{Bak66},
\citep[][\LJnow{p233}]{Wal82}
and
\citep[][\LJnow{p156}]{Hel75},
and published by Abd el-Mohsen Bakir.
As in all our three previous studies \citep{Por08,Jet13,Jet15},
we use {\it only the best preserved continuous calendar} 
which is found on pages recto III-XXX and
verso I-IX of papyrus Cairo 86637. 
The other texts and fragments contained in the 
same papyrus are ignored from this analysis 
because the connection of these fragments 
to the main calendar is not apparent and 
we do not know what year they describe, 
so combining any data points from these 
sources to the dataset created from the 
long Cairo Calendar would introduce 
a random noise component to the analysis.
All CC text passages we 
quote in this article 
have been translated by us from the hieroglyphic transcription
of \citet{Lei94}, % RefOK
assisted by the translations of
\citet[][in English]{Bak66} % RefOK
and \citet[][in German]{Lei94}. % RefOK

The synodic period of the Moon was discovered in CC with 
a statistical method called the Rayleigh test,
as well as a few other periods \citep[][\LJnow{p334}]{Por08}.  % RefOK
In a footnote of that study, 
it was noted that one seemingly less significant period, 
2.850 days, was rather close to the current 2.867 days period of
Algol ($\beta$ Persei). This star is a prototype of a class of
stars called eclipsing binaries.
The two stars, Algol~A and Algol~B,
orbit around a common centre of mass with a period
of 2.867 days.
Algol~A is brighter than Algol~B.
However, Algol~B has a larger
radius than Algol~A.
Our line of sight nearly coincides
with the orbital plane of this double star system.
Therefore, these stars eclipse each other during every orbital round.
In a primary eclipse, the dimmer Algol~B partly eclipses the brighter Algol~A.
This primary eclipse can be observed with naked eye.
In a secondary eclipse, the brighter Algol~A partly
eclipses the dimmer Algol~B,
but the decrease in total brightness of this binary system is so small 
that this secondary eclipse event can not be observed with naked eye.
Hence, the brightness of Algol appears to remain 
constant for a naked eye observer,
except during the primary eclipses.
These primary eclipses last about ten hours.
For most of the time,
Algol is brighter than its six close-by bright comparison stars
\citep[][their Figure 5a]{Jet13}.  % RefOK
During a primary eclipse, Algol first becomes dimmer for five hours
and then regains its brightness in another five hours.
For a few hours, Algol appears visibly dimmer than 
all its six comparison stars.
A naked eye observer can easily notice this as 
a clear change in Algol's constellation pattern.

The normalized Rayleigh test of the CC data 
confirmed the high significance of the 2.850 days period
\citep{Jet13}. % RefOK
The period increase from 2.850 to 2.867 days during the 
past three millennia gave a mass transfer 
rate estimate from Algol~B to Algol~A.
This estimate of \citet{Jet13}  % RefOK
agreed with the one predicted by the best 
evolutionary model of Algol
\citep[][\LJnow{p540}]{Sar93}.  % RefOK
A sequence of eight astronomical criteria was also presented
which proved that the ancient Egyptians could have discovered 
Algol's periodic variability with naked eyes 
\citep[][\LJnow{p9-10}]{Jet13},  % RefOK
\textit{i.e.} it is the star where it is easiest to discover regular 
short-term variability without the aid of a telescope. 
 
In the Hellenistic tradition, Algol was called ``the head of Gorgon''.
Similar tradition was continued in the Arabic name ``Demon's Head''. 
The name Algol is derived from the Arabic word, 
head of the Ghoul (ra's al-gh\={u}l) \citep{Dav57}.  % RefOK
These names seem to indicate that some exotic 
or foreboding feature or mutability 
was known in the folklore of the ancient peoples. 
All the way to medieval astrology, 
the ill omens associated with the ``evil eye'' of Algol 
were known, so it is actually surprising that 
it is so difficult to find any direct reference to Algol's 
variability in old astronomical texts \citep{Dav57}. 
The list of ill-omened names 
is so impressive \citep[][\LJnow{p332-333}]{All99} % RefOK
that it is unlikely that the variability 
would have gone undetected through millennia of practical star 
observing by the ancient Egyptians.

Of the modern astronomers,
Fabricius discovered the first variable star, Mira, in 1596.
The second variable star, Algol, was discovered by Montanari in 1669.
\citet{Goo83} determined the 2.867 days period of Algol 
in 1783. % RefOK
A close friend and tutor of John Goodricke, 
Edward Pigott, also discovered several 
new variable stars \citep{Hos79}.  % RefOK
In his last paper,  \citet[][\LJnow{p152}]{Pig05} 
argued that the brightness of Algol 
must have been constant in Antiquity, 
because the variability that he observed 
was so easy to notice with naked eyes. % RefOK
\citet[][\LJnow{p3}]{Kop46} suggested that those ancient discoveries 
``may have been buried in 
the ashes of the Library of Alexandria''. % RefOK
More recently, 
\citet{Wil00}
has presented the theory that classical mythology 
contains knowledge of the variability of various stars, 
including Algol. % RefOK
This star also seems to belong to the constellation 
called ``Elk'' by the Siberian shamans of the Khanty tribe, 
who have noticed that this animal sometimes 
loses one pair of legs \citep[][\LJnow{p58-65}]{Pen97}.  % RefOK

A statistical analysis of
28 selected words (hereafter SWs) of the mythological 
narratives of CC was performed
to find traces of the Egyptians' symbolism for Algol
\citep{Jet15}. % RefOK
We notate the SWs of that particular
study for example \SW{Horus} or \SW{Seth}
to distinguish them from other Egyptian deities such as Isis and Nephthys.
Out of all 28 SWs, 
the word \SW{Horus} had the strongest connection
to the 2.850 days periodicity \citep{Jet15}.  % RefOK
\SW{Horus}, etymologically ``the distant one'', 
was one of the earliest attested Egyptian deities. 
Predominantly a sky god or stellar god, 
the living king was identified as an 
earthly \SW{Horus} \citep[][\LJnow{p42-43}]{Roe94} and 
\citep[][\LJnow{p119-122}]{Mel01}.
Horus is described as a star in 
the oldest ancient Egyptian texts \citep{Kra16}.
Another deity, \SW{Seth}, the adversary of \SW{Horus}, 
was shown to be connected to the period of the Moon
\citep{Jet15}. % RefOK

Statistical analyses have 
confirmed the ancient Egyptian discovery 
of Algol's period \citep{Por08,Jet13,Jet15}. % RefOK
Here, our aim is to connect this astonishing ancient 
discovery to its contemporary cultural and historical background 
by presenting ten general arguments 
about CC (Sects \ref{hourstars}-\ref{astrophys}).
These arguments strongly support the idea
that the ancient Egyptian scribes had the possible
means and the motives to record Algol's period into CC.
The connection of CC mythological texts to the 
perceived behaviour of the Moon and Algol 
is verified in Sects \ref{Legends} and \ref{Reasoning}.

\section{Materials and Methods}

\subsection{Materials \label{materials}}

We use statistical methods to discover the principles of describing 
celestial phenomena in CC, thus no other Egyptian 
texts are used as material 
in the core analysis.
We begin with a general description of CC.

This document is one of the texts known 
as Calendars of Lucky and Unlucky Days.
In these Calendars the days of the year 
are assigned good and bad prognoses. 
Nine full and partial Calendars of Lucky and Unlucky Days 
have been discovered 
\citep[][\LJnow{p1-2}]{Lei94},
\citep[][\LJnow{p140-143}]{Tro89}
and
\citep[][\LJnow{p156}]{Hel75}.  % RefOK
Eight of them date to the New Kingdom, ca. 1550-1069 B.C., 
while one of them is from the Middle Kingdom, 
ca. 2030-1640 B.C. 
Papyrus Cairo 86637, the source of CC, was originally 
dated to the ninth regnal year of Ramses II \citep{Bru70},  % RefOK
around 1271-1270 B.C. according to the generally 
accepted chronology \citep[][\LJnow{p480-490}]{Sha00} % RefOK
which has been disputed \citep{Hub11}. % RefOK
However, the date is nowhere to be explicitly 
found \citep[][\LJnow{p1-2}]{Lei94}.  % RefOK
\citet[][\LJnow{p233}]{Wal82}   % RefOK
revised the date of the papyrus 
to the early 20th dynasty, 
around 1185-1176 B.C. 
We have also checked the paleographical 
correspondences of plentifully recurring signs, 
such as F35, G17, N5, O1
and R8 \citep[][\LJnow{p118,129,194,246,274}]{Wim95}, % RefOK
and these seem to support the conclusion of 
dating the manuscript to the latter half of the 19th dynasty 
or the beginning of the 20th, \textit{i.e.} 1244-1163 B.C. 
A compromise date 1224 B.C. was used in the astrophysical 
and astronomical computations \citep[][\LJnow{p1}]{Jet13},  % RefOK
as well as in the SW analysis \citep[][\LJnow{p1}]{Jet15}. % RefOK
The results of both of those studies
did not depend on the exact dating of CC.

CC is a calendar for the entire year. 
We use the daily prognoses of CC published in Table 1 of 
\citet{Jet13},  % RefOK
where 
the German notations by \citet[][\LJnow{p480-482}]{Lei94}  % RefOK
were used
(G=Gut= ``good'', S=Schlecht=``bad'').
CC is based on the Civil Calendar of 12 months of 
30 days each plus five additional epagomenal days 
for which no prognoses are given. The months were arranged 
into three seasons of four months each. These seasons were 
Akhet (flood season)\Egy, Peret (winter season)\Egy 
and Shemu (harvest season)\Egy. 
The conventionally given format for 
a calendar date is for example I Akhet 27 
for the 27th day of the first month of the Akhet season.

The CC texts systematically give a date, inscribed in red colour, 
and then three prognoses for that date (FAQ \ref{deitydate})
\footnote{Some frequently asked questions (FAQ) about our research have been
collected into Appendix B, where we give short answers those questions,
as well as indicate the sections of this manuscript
where the more detailed answers can be found.}. 
For example, the GGG prognosis 
combination for the date I Akhet 27 means that
all the three parts of the day are lucky. 
This fully positive prognosis is the most common for any day. 
\citet{Kem91}  % RefOK
noted that the ratio of good and bad prognoses 
in CC is close to the value of the so-called Golden Section, 
in accordance with modern psychological experiments 
regarding positive and negative judgements.

Generally speaking, on SSS days people 
were under a special threat 
to suffer from hunger, thirst and various illnesses. 
The prognoses of such days 
were attributed to mostly negative mythological events 
and children born on such a date 
might have been foretold to die of illness. 
On the other hand, 
those born on GGG days would live a long life. 
Such days were in general 
supposed to consist of joy, success, freedom, health and various feasts. 
While on SSS days some restrictions 
were suggested on journeying and consumption of foods, 
on GGG days it was recommended to give offerings and feasts 
to the gods \citep[][\LJnow{p138}]{Tro89}. % RefOK

In the longer and better preserved texts,
especially in CC,
there are descriptions of mythological events relating to the date 
and also some instructions on suggested behaviour during the day
(FAQ \ref{deitydate}). 
For example, regarding the day I Akhet 27 the description 
in CC, page recto VIII, 
reads that the god \SW{Horus} and his enemy \SW{Seth} 
are resting from their perpetual struggle. 
It is recommended not to kill any 
``snakes''\Egy during the day. 
The practical influence of the Calendars of Lucky and Unlucky Days 
on the life of ancient Egyptians is not exactly known. 
The various instructions and restrictions 
such as ``make offering to the gods of your city'' 
\citep[][\LJnow{p82}]{Lei94}  % RefOK
or ``do not go out of your house to any road on this day''
\citep[][\LJnow{p238}]{Lei94} % RefOK
seem to be presented in the context of the everyday life of a worker. 
It was suggested that the Calendars of Lucky and Unlucky Days would have 
determined the rest days for the 
workers \citep[][\LJnow{p153-155}]{Hel75},  % RefOK
but no correlation of the Lucky and Unlucky Days 
was found with days of kings' ascensions to throne, 
official building works, battles, journeys, court trials or working days, 
except when the day was also a regular feast date 
\citep[][\LJnow{p87-94}]{Dre72}.  % RefOK
In CC,  the prognosis of the first day of each month is always 
GGG and the day is called 
``feast''\Egy.  
On the other hand, the prognosis of the 20th day of each month is always SSS.

In most cases, the prognosis 
is homogeneous for the whole day (\textit{i.e.} GGG or SSS). 
There are only 29 heterogeneous prognoses in CC. 
These days provide a glimpse into the logic behind the day division. 
Generally speaking, arrangement into morning, mid-day and evening 
is obvious, but these can be defined in multiple ways. 
For example, the prognosis for the date I Akhet 8 in CC is GGS. 
The text advises one not to go out during the 
 ``night''\Egy. 
The prognosis for I Akhet 25 is also GGS 
but the text advises one not to go out during the 
 ``evening''\Egy. 
Thus it remains unclear if the third part of 
the day comprises night hours as well. 
\citet[][\LJnow{p2-7}]{Jet13} % RefOK
showed that the period analysis results
for CC did not depend on how the three prognoses were distributed
within each day.

The practice of assigning good and bad omens to days of the year 
seems rather close to astrology and reading predictions from the stars, 
and indeed the Calendar of Lucky and Unlucky Days 
was mixed with Babylonian based astrology in the Greek and 
Roman times \citep[][\LJnow{p38-55}]{Lei95}. % RefOK
But it is to be noted that celestial matters 
did not fully determine 
the prognoses in the
Calendar of Lucky and Unlucky days, 
but played a part in it alongside natural cycles 
such as the floods of the Nile, 
or the seasonal dangers presented by winds, wild animals and illnesses. 
There is also plenty of evidence for various kinds of ritual recurrence, 
such as that III Akhet 26 
is described the 
strengthening of ``the djed-pillar''\Egy ~
\citep[][a ritual object whose 
raising is connected to the myth of the resurrection of the god Osiris
who was killed by Seth]{Hel75}
and 
II Peret 6 is described 
the erection of ``the djed-pillar'', 
with a separation of exactly 70 days, 
the approximate interval between 
Sirius' heliacal setting and Sirius' heliacal rising. 
It was considered the ideal duration of funerary ceremonies 
because the star was believed to spend this time in the underworld, 
undergoing rituals of purification. 
CC also makes explicit references to the heliacal rising, 
culmination and heliacal setting of certain 
hour-stars \citep{Har03}. % RefOK
All prognoses based on this type of aperiodic events 
induce statistical noise into CC. 
When applying period analysis to the CC data,
this noise interferes with the detection of any periodic signal.

When searching for regular periodic astronomical phenomena
in CC, one should realize that
only a few events relating to celestial objects, 
however important they were considered to be, 
could have determined 
an extensive and significant
set of periodic prognoses. 
For example, the heliacal rising of a star is a yearly event, and
may affect the prognosis of one day. 
Thus, it can not be discovered from the calendars by period analysis. 
The synodic periods of planets, 
because of their length, are also out of the question
\citep[][\LJnow{p13}]{Jet13}.  % RefOK
Except for the Moon,
the only other detectable astronomical 
periods in CC
are those of the regular brightness changes of 
a variable star \citep[][\LJnow{p9}]{Jet13}.  % RefOK

\subsection{Methods \label{methods}}
\begin{table}[!t]
\centering
\caption{{\bf ``GGG'' prognosis texts mentioning
Horus, Wedjat or Sakhmet.}
The columns are 
SW (Selected word),
ancient Egyptian month (``Month''),
day ($D$), numerical month value ($M$), time point ($g(D,M)$)
and the phase angles
($\Theta_{\mathrm{Algol}}$ and $\Theta_{\mathrm{Moon}}$).
All values are in the order of increasing $\Theta_{\mathrm{Algol}}$,
because this allows an easy comparison with the results
shown in List 1 and Figure \ref{figone}.}
\begin{tabularx}{0.485\textwidth}{p{1cm}P{1.2cm}YYP{1cm}YP{0.7cm}}
SW & Month & $D$ & $M$ & $g(D,M)$ & $\Theta_{\mathrm{Algol}}$ &  $\Theta_{\mathrm{Moon}}$ \\
\hline
     Horus &   II Akhet &  14 &   2 &    43.33 &    6 &  124 \\

    Wedjat &    I Peret &   1 &   5 &   120.33 &   13 &  341 \\

   Sakhmet &    I Peret &   1 &   5 &   120.33 &   13 &  341 \\

     Horus &   IV Shemu &  19 &  12 &   348.33 &   13 &  234 \\

     Horus &    I Akhet &  27 &   1 &    26.33 &   19 &  278 \\

     Horus &  III Akhet &  24 &   3 &    83.33 &   19 &  251 \\

     Horus &  III Peret &   1 &   7 &   180.33 &   32 &  351 \\

     Horus &  III Akhet &  27 &   3 &    86.33 &   38 &  287 \\

     Horus &  III Shemu &  15 &  11 &   314.33 &   38 &  180 \\

     Horus &    I Shemu &   1 &   9 &   240.33 &   51 &    0 \\

    Wedjat &   II Akhet &   3 &   2 &    32.33 &   57 &  351 \\

     Horus &    I Shemu &   7 &   9 &   246.33 &   88 &   73 \\

     Horus &  III Akhet &  28 &   3 &    87.33 &  164 &  300 \\

     Horus &   II Shemu &   1 &  10 &   270.33 &  240 &    5 \\

   Sakhmet &   IV Akhet &  16 &   4 &   105.33 &  278 &  158 \\

     Horus &  III Peret &  23 &   7 &   202.33 &  291 &  258 \\

     Horus &  III Akhet &  29 &   3 &    88.33 &  291 &  312 \\

   Sakhmet &    I Peret &   9 &   5 &   128.33 &  303 &   78 \\

    Wedjat &   II Shemu &  30 &  10 &   299.33 &  303 &  358 \\

   Sakhmet &    I Peret &  29 &   5 &   148.33 &  309 &  321 \\

     Horus &    I Akhet &  18 &   1 &    17.33 &  322 &  168 \\

    Wedjat &    I Shemu &   6 &   9 &   245.33 &  322 &   61 \\
\end{tabularx}
\label{tableone}%tab1
\end{table}
\begin{table}[!t]
\centering
\caption{ {\bf ``SSS'' prognosis texts mentioning Horus, Wedjat or
Sakhmet.} Notations are as in Table \ref{tableone}.}
\begin{tabularx}{0.485\textwidth}{p{1cm}P{1.2cm}YYP{1cm}YP{0.7cm}}
SW & Month & $D$ & $M$ & $s(D,M)$ & $\Theta_{\mathrm{Algol}}$ &  $\Theta_{\mathrm{Moon}}$ \\
\hline
     Horus &   IV Peret &   5 &   8 &   214.33 &    6 &   44 \\

     Horus &    I Akhet &  26 &   1 &    25.33 &  253 &  265 \\

     Horus &  III Shemu &  11 &  11 &   310.33 &  253 &  132 \\

    Wedjat &   II Peret &  10 &   6 &   159.33 &  259 &   95 \\

   Sakhmet &   IV Peret &  27 &   8 &   236.33 &  265 &  312 \\

   Sakhmet &   II Peret &  13 &   6 &   162.33 &  278 &  132 \\

   Sakhmet &   II Shemu &   7 &  10 &   276.33 &  278 &   78 \\

     Horus &    I Shemu &  20 &   9 &   259.33 &  291 &  231 \\
\end{tabularx}
\label{tabletwo}%tab2
\end{table}
\begin{table}[!t]
\centering
\caption{{\bf ``GGG'' prognosis texts mentioning Horus, Seth or Osiris.}
Notations are as in Table \ref{tableone}, except that
all values are in the order of increasing $\Theta_{\mathrm{Moon}}$.}
\begin{tabularx}{0.485\textwidth}{p{1cm}P{1.2cm}YYP{1cm}YP{0.7cm}}
SW & Month & $D$ & $M$ & $g(D,M)$ & $\Theta_{\mathrm{Algol}}$ &  $\Theta_{\mathrm{Moon}}$ \\
\hline
     Horus &    I Shemu &   1 &   9 &   240.33 &   51 &    0 \\

     Horus &   II Shemu &   1 &  10 &   270.33 &  240 &    5 \\

    Osiris &   II Shemu &   1 &  10 &   270.33 &  240 &    5 \\

    Osiris &  III Peret &   6 &   7 &   185.33 &  303 &   51 \\

      Seth &   IV Akhet &   9 &   4 &    98.33 &  114 &   73 \\

     Horus &    I Shemu &   7 &   9 &   246.33 &   88 &   73 \\

    Osiris &   IV Akhet &  11 &   4 &   100.33 &    6 &   98 \\

     Horus &   II Akhet &  14 &   2 &    43.33 &    6 &  124 \\

    Osiris &   II Akhet &  16 &   2 &    45.33 &  259 &  149 \\

      Seth &   IV Shemu &  13 &  12 &   342.33 &  335 &  161 \\

    Osiris &   IV Shemu &  13 &  12 &   342.33 &  335 &  161 \\

     Horus &    I Akhet &  18 &   1 &    17.33 &  322 &  168 \\

     Horus &  III Shemu &  15 &  11 &   314.33 &   38 &  180 \\

    Osiris &   II Peret &  17 &   6 &   166.33 &   63 &  180 \\

     Horus &   IV Shemu &  19 &  12 &   348.33 &   13 &  234 \\

     Horus &  III Akhet &  24 &   3 &    83.33 &   19 &  251 \\

     Horus &  III Peret &  23 &   7 &   202.33 &  291 &  258 \\

     Horus &    I Akhet &  27 &   1 &    26.33 &   19 &  278 \\

      Seth &    I Akhet &  27 &   1 &    26.33 &   19 &  278 \\

     Horus &  III Akhet &  27 &   3 &    86.33 &   38 &  287 \\

      Seth &  III Akhet &  27 &   3 &    86.33 &   38 &  287 \\

     Horus &  III Akhet &  28 &   3 &    87.33 &  164 &  300 \\

    Osiris &  III Akhet &  28 &   3 &    87.33 &  164 &  300 \\

     Horus &  III Akhet &  29 &   3 &    88.33 &  291 &  312 \\

      Seth &  III Akhet &  29 &   3 &    88.33 &  291 &  312 \\

    Osiris &  III Peret &  28 &   7 &   207.33 &  202 &  319 \\

     Horus &  III Peret &   1 &   7 &   180.33 &   32 &  351 \\
\end{tabularx}%tab3
\label{tablethree}
\end{table}

We relate CC texts to astronomical events 
by the phase angles calculated from the days
that the texts refer to (Eqs. \ref{edays}-\ref{phaseangles}).
We select four samples from CC (Tables \ref{tableone}-\ref{tablefour})
which give us two lists of CC text passages (Lists 1 and 2).

Recently, a statistical study was made of
the occurrence of 28 different SWs in the CC prognosis 
texts \citep[][\LJnow{p3}]{Jet15}. % RefOK
The occurrences of individual SWs were studied separately.
The lucky prognosis texts mentioning \SW{Horus} were studied 
in greater detail, and a few unlucky texts mentioning 
\SW{Sakhmet} or \SW{Seth}.
Those texts were taken as such from the CC translation of 
\citet{Bak66}. % RefOK

We downloaded these SW data \citep[][\LJnow{p1}]{Jet15}  % RefOK
from the Dryad database\footnote{{\Mydryad}},
where the respective ASCII file-name is {\it data2.txt}.
This gave us the dates when any particular SW is mentioned in CC.
Here, we concentrate on the following five particular SWs: 
\SW{Horus}, \SW{Wedjat}, \SW{Sakhmet}, \SW{Seth} and \SW{Osiris}.
These five deities are the most relevant ones regarding the two 
prominent myths ``The Destruction of Mankind'' and 
``The Contendings of Horus and Seth'' that will be described in
Sect. \ref{Legends}.

\SW{Horus}, etymologically the distant one, 
was a sky god or stellar god associated with kingship and order. 
\citet[][\LJnow{p137-141}]{Kra16}
suggests that Horus was originally a stellar god who later 
became subordinated to solar mythology. 
Already the earliest texts regarding Horus describe him as 
the ``Foremost star of the sky''. In the Pyramid Texts, 
a younger Horus is called Horus-son-of-Isis and is distinguished 
from the elder Horus (Haroeris). 
According to Krauss these would be Venus as the morning star 
and the evening star. On one hand the connection of the planet Venus, 
usually considered feminine, with the king would be unique to Egypt. 
Yet Venus was certainly associated with Benu, 
the divinity who in the creation myth laid the first stone 
{\it benben},
which became Earth, upon the primal sea. 
Horus' rival god Seth was the embodiment of disorder, 
identified in some sources with the planet Mercury, 
the other inner planet besides Venus. 
In the most commonly known mythological narrative Osiris, 
the father of Horus, was killed by Seth. 
Horus avenged his father and the resurrected Osiris 
was to be considered the patron of the dead, 
especially the dead king 
\citep[][\LJnow{p119-120}]{Mel01}.
Sakhmet was a lion goddess associated with the scorching, 
destructive power of the Sun, also called Wadjet
or the Eye of Horus 
\citep[][\LJnow{p361}]{Lei03}
and \citep[][\LJnow{p512-513}]{Jon01}.

While\begin{table}[!t]
\centering
\caption{{\bf ``SSS'' prognosis texts mentioning Horus, Seth or Osiris.}
Notations are as in Table \ref{tablethree}.}
\begin{tabularx}{0.485\textwidth}{p{1cm}P{1.2cm}YYP{1cm}YP{0.7cm}}
SW & Month & $D$ & $M$ & $s(D,M)$ & $\Theta_{\mathrm{Algol}}$ &  $\Theta_{\mathrm{Moon}}$ \\
\hline
     Horus &   IV Peret &   5 &   8 &   214.33 &    6 &   44 \\

      Seth &   II Akhet &  12 &   2 &    41.33 &  114 &  100 \\

      Seth &  III Akhet &  13 &   3 &    72.33 &   69 &  117 \\

    Osiris &  III Akhet &  13 &   3 &    72.33 &   69 &  117 \\

    Osiris &  III Akhet &  14 &   3 &    73.33 &  196 &  129 \\

     Horus &  III Shemu &  11 &  11 &   310.33 &  253 &  132 \\

      Seth &   IV Shemu &  11 &  12 &   340.33 &   82 &  137 \\

    Osiris &    I Peret &  14 &   5 &   133.33 &  215 &  139 \\

      Seth &  III Akhet &  18 &   3 &    77.33 &  341 &  178 \\

      Seth &  III Peret &  17 &   7 &   196.33 &  253 &  185 \\

      Seth &   IV Peret &  17 &   8 &   226.33 &   82 &  190 \\

    Osiris &   IV Akhet &  19 &   4 &   108.33 &  297 &  195 \\

      Seth &   II Akhet &  20 &   2 &    49.33 &   44 &  197 \\

     Horus &    I Shemu &  20 &   9 &   259.33 &  291 &  231 \\

     Horus &    I Akhet &  26 &   1 &    25.33 &  253 &  266 \\

      Seth &    I Akhet &  26 &   1 &    25.33 &  253 &  266 \\

      Seth &   IV Peret &  24 &   8 &   233.33 &  246 &  275 \\
\end{tabularx}
\label{tablefour}%tab4
\end{table} \citet{Jet15}  % RefOK
studied {\it some}
texts mentioning \SW{Horus}, \SW{Sakhmet} or \SW{Seth}, 
and used the text from the CC translation of \citet{Bak66}, % RefOK
we study {\it all} prognosis texts mentioning 
{\it any} of the above mentioned five deities, 
and we use our 
{\it own}
translations of these CC texts.

We calculate the ``Egyptian days'' for these SWs 
from
\begin{align}
N_{\mathrm{E}}=30(M-1)+D,
\label{edays}
\end{align}

\noindent
where $M$ is the month
and $D$ is the day of the date in CC 
\citep[][Table 1]{Jet13}. % RefOK
The SW dates are transformed \citep{Jet15} % RefOK
into time points with the relation
\begin{align}
t=t(D,M)=N_{\mathrm{E}}-1+0.33.
\label{timepoints}
\end{align}
We use the notations $g=g(D,M)=t(D,M)$ and $s=s(D,M)=t(D,M)$
for the time points of GGG and SSS prognosis days,
because these two prognosis samples were analysed 
and studied separately \citep{Jet13,Jet15}. % RefOK

For any period value $P$,
the phases of $t$ are 
\begin{align}
\phi=\mathrm{FRAC}[(t-t_0)/P],
\label{vaiheet}
\end{align}
where $\mathrm{FRAC}$ removes the integer part of $(t-t_0)/P$
and $t_0$ is the zero epoch. 
In other words, $\mathrm{FRAC}$ removes the number of 
full $P$ rounds completed after the zero epoch $t_0$.
The phase angles are 
\begin{align}
\Theta= 360\aste \phi.
\label{phaseangles}
\end{align}
\citet{Jet13}  % RefOK
discovered two significant periods,
$P_{\mathrm{Algol}}=2.850$
and 
$P_{\mathrm{Moon}}=29.6$ days,
in the lucky prognoses of CC.
In their next study,
they determined these two 
ephemerides \citep{Jet15} for the phases of Eq. \ref{vaiheet}
\begin{align}
t_0=0.53, & P=P_{\mathrm{Algol}}=2.850 \label{Aephe} \\
t_0=3.50, & P=P_{\mathrm{Moon}}=29.6  \label{Mephe} 
\end{align}
The phase angles $\Theta$ (Eq. \ref{phaseangles})
computed with the ephemerides of 
Eqs. \ref{Aephe} and \ref{Mephe} are hereafter denoted with
$\Theta_{\mathrm{Algol}}$ and $\Theta_{\mathrm{Moon}}$,
respectively. 
We also use their eight abbreviations \citep[][\LJnow{p6-7}]{Jet15} % RefOK
\begin{itemize}
\item[Aa] $\equiv \Theta_{\mathrm{Algol}}=0\aste$ 
= Mid-epoch of Algol's secondary eclipse
\item[Ab] $\equiv \Theta_{\mathrm{Algol}}=90\aste$ 
= Between mid-epochs of secondary and primary eclipse
\item[Ac] $\equiv \Theta_{\mathrm{Algol}}=180\aste$
= Mid-epoch of Algol's primary eclipse
\item[Ad] $\equiv \Theta_{\mathrm{Algol}}=270\aste$
= Between mid-epochs of primary and secondary eclipse
\item[Ma] $\equiv \Theta_{\mathrm{Moon}}=0\aste$ 
= Full Moon
\item[Mb] $\equiv \Theta_{\mathrm{Moon}}=90\aste$ 
= Between Full and New Moon
\item[Mc] $\equiv \Theta_{\mathrm{Moon}}=180\aste$
=  New Moon
\item[Md] $\equiv \Theta_{\mathrm{Moon}}=270\aste$
= Between New and Full Moon
\end{itemize}

All $D$, $M$, $g(D,M)$, $s(D,M)$, $\Theta_{\mathrm{Moon}}$ and
$\Theta_{\mathrm{Algol}}$ values of \SW{Horus}, \SW{Wedjat},
\SW{Sakhmet}, \SW{Seth} and \SW{Osiris} are given
in Tables \ref{tableone}-\ref{tablefour}.

We study all CC passages mentioning 
\SW{Horus},
\SW{Wedjat},
\SW{Sakhmet},
\SW{Seth}
and
\SW{Osiris}.
These passages give the date first, inscribed in red colour. 
Then follow the daily prognoses, and the descriptive prognosis text. 
The time point for every date is unambiguous,
because the structure of CC is regular, $12 \times 30$ days 
(Eqs. \ref{edays} and \ref{timepoints}).
Hence, the time points for the prognoses, the SWs and 
the prognosis texts describing the actions of deities
are also unambiguous. 
\footnote{
The calendar dates are fixed and known even if the texts or the deities 
mentioned in these texts were not studied at all.
The deities mentioned in these texts
do not determine the dates (FAQ \ref{deitydate}).}
The exact dating of the CC, 
as a historical document, 
is irrelevant in the current analysis, 
like it also was in the previous statistical 
studies \citep{Jet13,Jet15}. % RefOK
Adding any arbitrary constant to the time points of 
Eq. \ref{timepoints} shifts all phase angles $\Theta$ with the same amount,
\textit{e.g.} the $\Theta_{\mathrm{Algol}}$ values of all passages of List 1.
This is the reason why our results based on Lists 1 and 2
do not depend on such shifts (FAQ \ref{dating}).

The number of passages mentioning some SWs is very small
(\textit{e.g.} $n=3$ for \SW{Sakhmet} in Table \ref{tabletwo}).
It is therefore necessary to explain 
how can we draw reliable statistical conclusions
from the analysis of such data (FAQ \ref{smallsamples}).
Firstly, the periods $P_{\mathrm{Algol}}$ and $P_{\mathrm{Moon}}$
were detected from large samples of over five hundred time
points and these periodicities were extremely 
significant \citep{Jet13}. % RefOK
For example, the period $P_{\mathrm{Algol}}$ reached critical
levels $Q^{*} < 0.0001$ \citep[][Table 7]{Jet13},
\textit{i.e.} the probability for this period being real was $1-Q^{*} >0.9999$.
Secondly, the ephemerides of Eqs. \ref{Aephe} and \ref{Mephe} 
are also very reliable,
because they were determined from the same large data samples
\citep{Jet15}.
Thirdly, although the Rayleigh test significance estimates
computed by {\citet[][their Eq. 8: $Q_z$]{Jet15}} for some
smaller samples were not reliable, the binomial 
distribution significance estimates for the very same samples
were certainly reliable (their Eq. 13: $Q_{\mathrm{B}}$).
Fourthly, the {\it order} of the passages in Lists 1 and 2 is
the same (\textit{i.e.} unambiguous) for any arbitrary epoch $t_0$
in  Eqs. \ref{Aephe} and \ref{Mephe}. 
For these four reasons,
the phase angles computed from the ephemerides 
of Eqs. \ref{Aephe} and \ref{Mephe}
can be used just like the time given by an accurate modern clock.
For example, such a clock shows that most people go to sleep before midnight.
It is irrelevant if only a few (small $n$), 
or many (large $n$), people go to sleep. 
Rearranging the texts of CC into the increasing order of $\Theta_{\mathrm{Algol}}$
may show what the authors of CC
wrote about \SW{Horus}
at different phases of the cycle 
(FAQ \ref{amiss}).

\begin{figure*}[!t]\capstart
\centering
\includegraphics[width=0.85\textwidth]{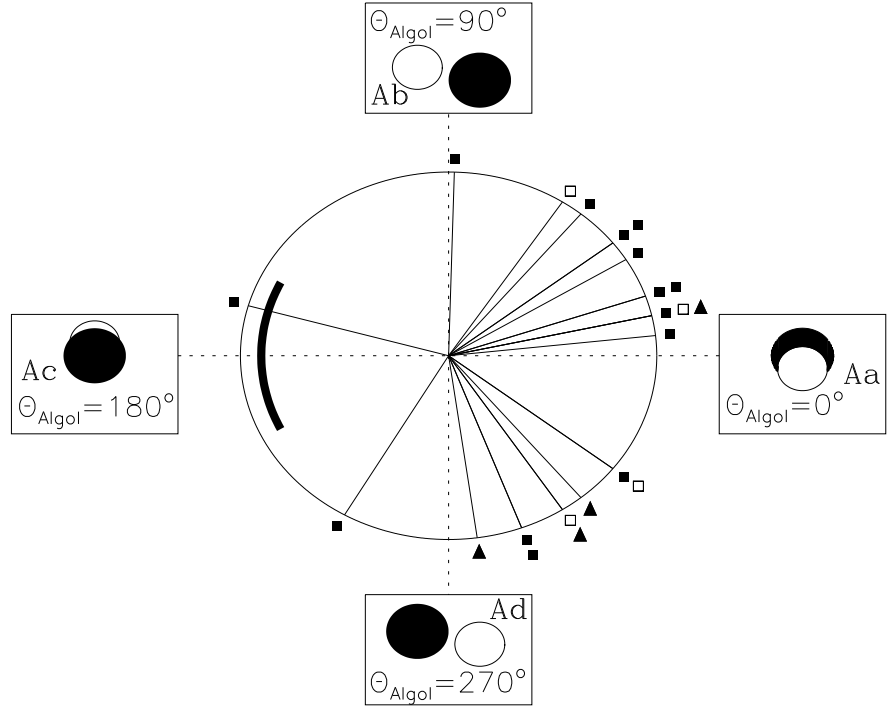}
\caption{{\bf $\Theta_{\mathrm{Algol}}$ phase angles
of lucky time points of Table \ref{tableone}.} 
Time runs in the counter-clockwise direction on this circle.
Epochs Aa, Ab, Ac and Ad are separated by 90 degrees
and they are denoted with dotted straight lines.
The
relative locations of Algol~A (white disk) and Algol~B (black disk)
at these four epochs are shown in the small boxes.
Primary and secondary eclipses of Algol occur at Ac and Aa, respectively.
The thick curved line centered at Ac outlines the phase angles of 
the 10 hour primary eclipse of Algol 
at $153.7\aste <$ $\Theta_{\mathrm{Algol}}<$ $206.3\aste$.
The phase angle values $\Theta_{\mathrm{Algol}}$  of \SW{Horus} (closed squares),
\SW{Wedjat} (open squares) and \SW{Sakhmet} (closed triangles)
are denoted with continuous straight lines. }
\label{figone}%fig1
\end{figure*}
\begin{figure*}[!t]\capstart
\centering
\includegraphics[width=0.85\textwidth]{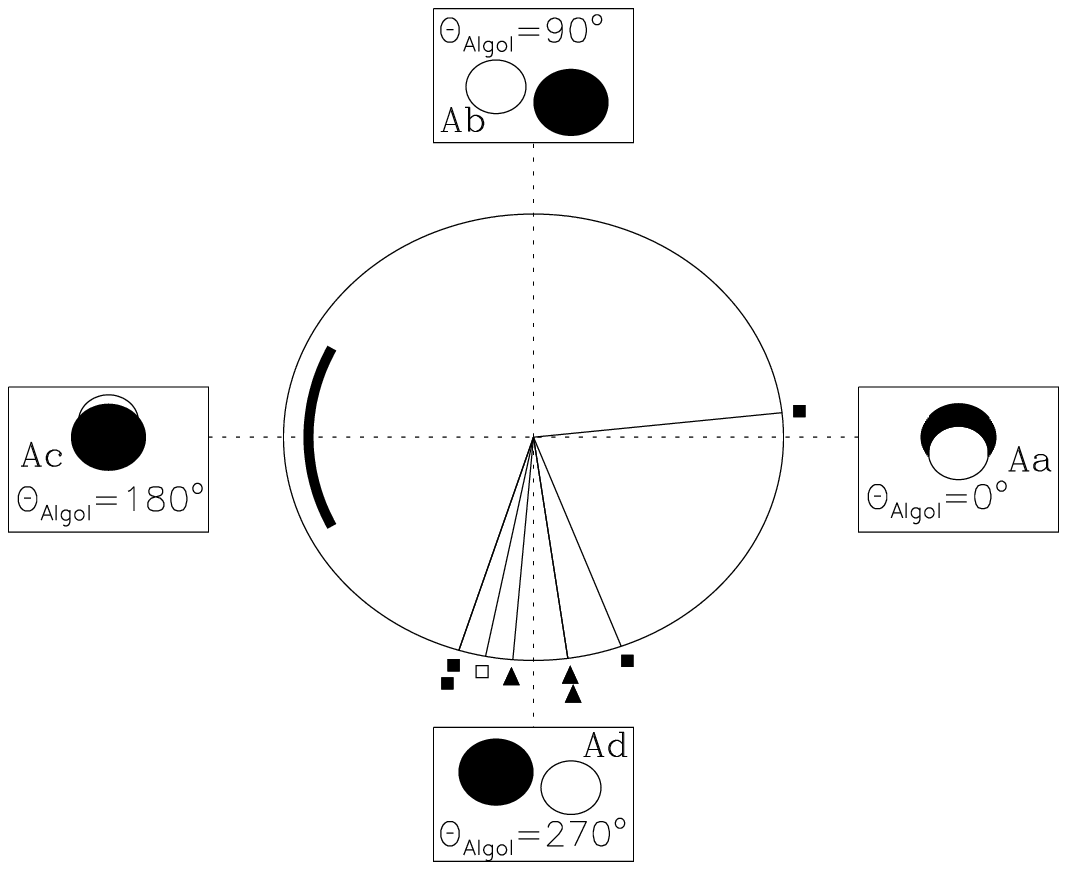}
\caption{{\bf $\Theta_{\mathrm{Algol}}$ of unlucky 
time points of Table \ref{tabletwo}.}
Notations as in Figure \ref{figone}.}
\label{figtwo}%fig2
\end{figure*}

It is possible to dispute our translations of the texts of Lists 1 and 2
(FAQ \ref{translation}), 
but these translations are relevant only for the two Arguments VII and VIII,
\textit{i.e.} the validity of the eight other arguments 
in Sects. \ref{hourstars}-\ref{astrophys} does not depend on
these translations.
In these translations,
we have used the English translation by \citet{Bak66}
and the German translation by \citet{Lei94},
and his transcript of the original papyrus,
as well as photos of the original papyrus.
For example, all 460 SW identifications by 
\citet{Jet15} 
and \citet{Lei94} were identical, and here we use the same SW list. 
Some words or sentences could be translated
differently, but that would not change the general description of the 
course of events in the translated passages of Lists 1 and 2. 
Also the prognoses, which were taken as such from \citet{Lei94}, 
are independent of any translation nuances.

A non-parametric method, the Rayleigh test, has been applied
to the series of $n$ time points $t_1, t_2, ... t_n$ of 
CC \citep{Jet13,Jet15}.
Here we study the SWs and the CC texts of these time points.
These time points are circular data and a ``non-parametric'' 
method means that there is no model. 
It has been suggested that we should apply a $\chi^2$-test to our data
(FAQ \ref{chitest}).
This ``parametric'' test could be applied if the format of our data were
$y(t_1), y(t_2), ... y(t_n)$, \textit{i.e.} a time series, like magnitudes
of a star as function of time.
The value of this test statistic is
$\chi^2= \sum_i^n [y(t_i)-g(t_i)]^2/\sigma_i^2 ,$
where $g(t_i)$ is the value of the model at $t_i$ and
$\sigma_i$ is the error of $y(t_i)$. 
However, we can not apply this $\chi^2$-test, 
because we have no time series, no model and no errors.

\section{Results  \label{results}}

\subsection{Algol: 
Horus, Wedjat and Sakhmet passages of List 1}

The lucky and unlucky days of CC texts mentioning
\SW{{Horus}}, \SW{Wedjat} or \SW{Sakhmet} 
are given in Tables \ref{tableone} and \ref{tabletwo}, respectively.
We rearrange our translations of
CC passages of \SW{Horus}, \SW{Wedjat} and
\SW{Sakhmet} into the order of increasing $\Theta_{\mathrm{Algol}}$ in
List 1 of Appendix C.
These passages are discussed in Sect. \ref{Reasoning}.
We highlight the unlucky CC prognosis
text passages with red colour, 
to visually distinguish these unlucky texts from the lucky ones, 
since red colour is also used in CC for writing the prognosis ``bad''
 and the name of the feared serpent creature Apep. 
Other texts in papyrus Cairo 86637 display an even 
more varied use of red colour for the sake of emphasis 
and captioning \citep[][\LJnow{p7}]{Bak66}.  % RefOK

The $\Theta_{\mathrm{Algol}}$ values of the $g(D,M)$ and $s(D,M)$ time points
of List 1 are shown separately in Figures \ref{figone} and \ref{figtwo}.
The relative positions of Algol A (white disk) and Algol B (black disk)
at points Aa, Ab, Ac and Ad are shown in four small boxes 
of Figures \ref{figone} and \ref{figtwo}.
For a naked eye observer, Algol's brightness appears constant,
except for the 10 hour dimming during $\Theta_{\mathrm{Algol}}$ values marked 
with a thick curved line centered at Ac.
These eclipse phase angles are in the interval 
$153.7\aste \le \Theta_{\mathrm{Algol}} \le 206.3\aste$.
Time runs in the counter-clockwise direction. 
One complete orbital round $P_{\mathrm{Algol}}$ is 
Aa $\rightarrow$ 
Ab $\rightarrow$ 
Ac $\rightarrow$ 
Ad $\rightarrow$ 
Aa.

The lucky time points $g(D,M)$ of SWs having 
$-90\aste < \Theta_{\mathrm{Algol}} < 90\aste$
amplify the $P_{\mathrm{Algol}}=2.^{\mathrm{d}}85$ signal \citep{Jet15}. % RefOK
The closer a $\Theta_{\mathrm{Algol}}$ value of some $g(D,M)$ is to 
the point Aa at $\Theta_{\mathrm{Algol}}=0\aste$ in Figure \ref{figone}, 
the greater is the amplifying impact of this $g(D,M)$ value
on the $P_{\mathrm{Algol}}$ signal.
A previous study by \citet{Jet15} showed 
that of all their 28 SWs, \SW{Horus} had the strongest impact 
on the $P_{\mathrm{Algol}}$ signal.
The other remaining SWs having 
an impact on the $P_{\mathrm{Algol}}$ 
signal were 
\SW{Re}, 
\SW{Wedjat}, 
\SW{Followers}, 
\SW{Sakhmet} and 
\SW{Ennead}.

\subsection{Moon: Horus, Seth and Osiris passages of List 2}
\begin{figure*}[!t]\capstart
\centering
\includegraphics[width=0.78\textwidth]{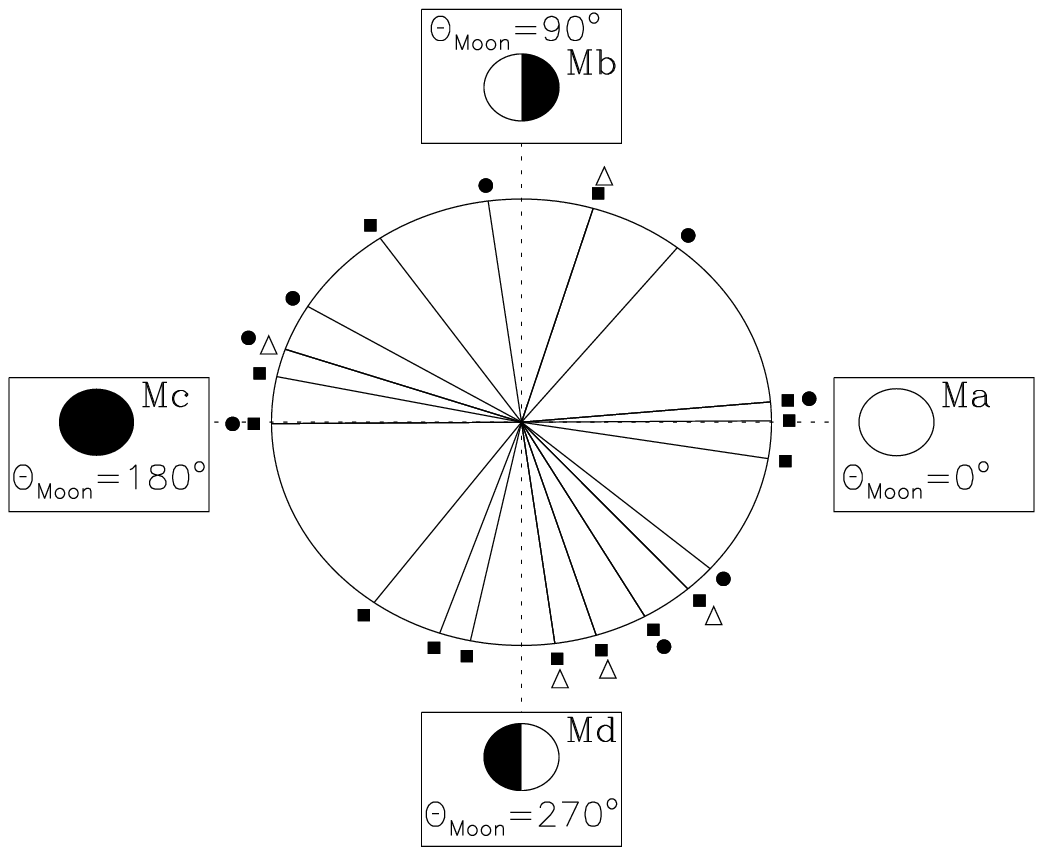}
\caption{
{\bf $\Theta_{\mathrm{Moon}}$ of the lucky time points of 
Table \ref{tablethree}.} 
Time runs in the counter-clockwise direction on this circle.
Epochs Ma, Mb, Mc and Md are separated by 90 degrees
and they are denoted with dotted straight lines.
The  
phases of the Moon are shown in the small boxes.
The Full and the New Moon occur at Ma and Mc, respectively.
The phase angle values $\Theta_{\mathrm{Moon}}$  of 
\SW{Horus} (closed squares)
\SW{Seth} (open triangles) and
\SW{Osiris} (closed circles) are denoted with continuous straight lines.
}
\label{figthree}
\end{figure*}
\begin{figure*}[!t]\capstart
\centering
\includegraphics[width=0.78\textwidth]{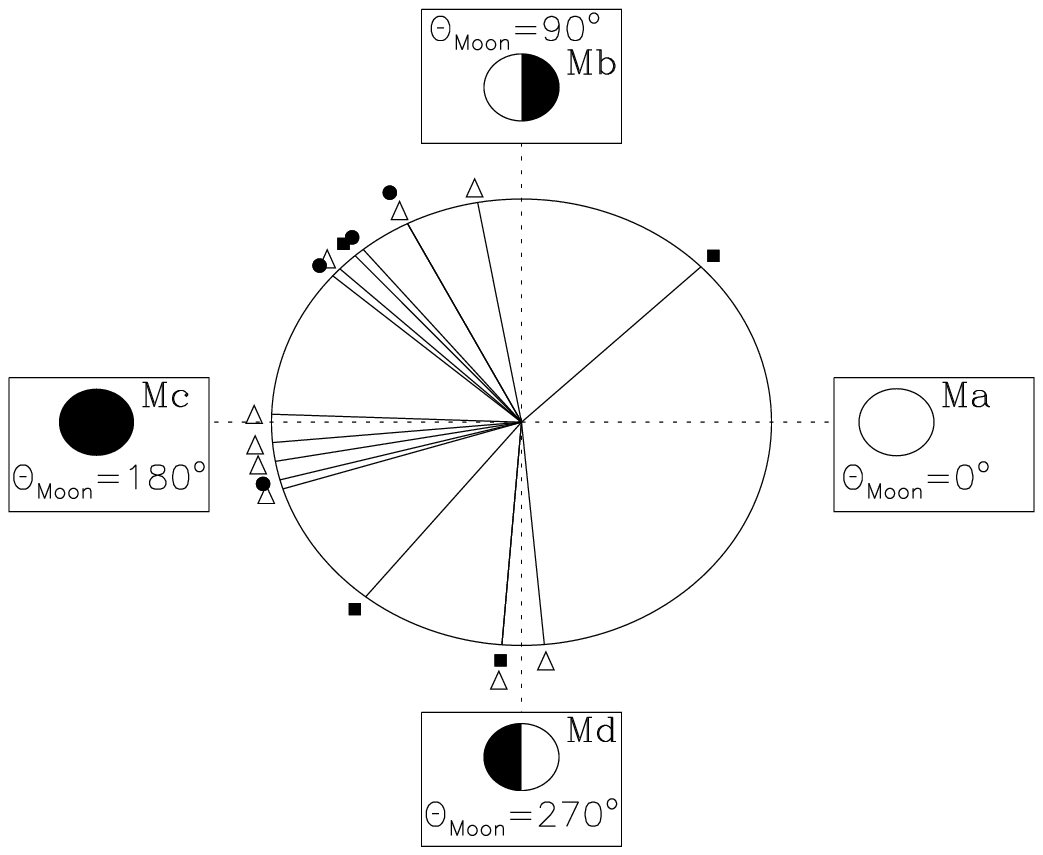}
\caption{{\bf $\Theta_{\mathrm{Moon}}$ of the unlucky time points of Table \ref{tablefour}.}
Notations as in Figure \ref{figthree}.}
\label{figfour}
\end{figure*}

The lucky and unlucky days of CC texts mentioning \SW{Horus},
\SW{Seth} or \SW{Osiris} 
are given in Tables \ref{tablethree} and \ref{tablefour}.
Our translations of CC passages mentioning 
\SW{Horus}, \SW{Seth} and
\SW{Osiris} are rearranged
into the order of increasing $\Theta_{\mathrm{Moon}}$
in List 2 of Appendix D.
We discuss these passages in Sect. \ref{Reasoning}.

The $\Theta_{\mathrm{Moon}}$ values of the $g(D,M)$ and  $s(D,M)$
time points of List 2 are shown in Figs \ref{figthree} and \ref{figfour}.
The appearance of the  lunar disk at points Ma, Mb, Mc and Md
is illustrated in four small boxes 
of Figures \ref{figthree} and \ref{figfour}.
Again, time runs in the counter-clockwise direction, 
where one complete synodic lunar month $P_{\mathrm{Moon}}$ is 
Ma $\rightarrow$ 
Mb $\rightarrow$ 
Mc $\rightarrow$ 
Md $\rightarrow$ 
Ma.

The time points 
$g(D,M)$ of SWs with phase angles $\Theta_{\mathrm{Moon}}$
close to $\Theta_{\mathrm{Moon}}=0\aste$ 
amplify the $P_{\mathrm{Moon}}=29.^{\mathrm{d}}6$ signal.
\citet{Jet15} % RefOK
showed that
of all their 28 SWs, \SW{Earth} and \SW{Heaven}
had the strongest impact on the $P_{\mathrm{Moon}}$ signal,
\textit{i.e.} their lucky prognoses were close to the Ma point. 
This is natural because lunar feast dates 
where often described as feasts 
in \SW{Earth} and in \SW{Heaven} (FAQ \ref{amiss}).
The other SWs connected to $P_{\mathrm{Moon}}$ were
\SW{Busiris}, 
\SW{Rebel}, 
\SW{Thoth} and 
\SW{Onnophris}
\citep{Jet15}. % RefOK
The unlucky time points $s(D,M)$ of two SWs, \SW{Seth} and \SW{Osiris},
pointed to the opposite direction, $\Theta_{\mathrm{Moon}}=180\aste$.
The CC texts of \SW{Seth} strongly indicated that  
$\Theta_{\mathrm{Moon}}=180\aste$ coincided with the New Moon.
Hence, it was 
concluded that $\Theta_{\mathrm{Moon}}=0\aste\equiv$ Ma 
represented the Full Moon \citep{Jet15}. % RefOK
These connections are hardly surprising either because \SW{Osiris}, 
also called \SW{Onnophris}, 
was identified with the Moon during the New Kingdom
\citep[][\LJnow{p480-482}]{Kap01}. % RefOK
\SW{Thoth} was another known lunar god \citep{Lei03}.
\SW{Busiris} was the place of origin for \SW{Osiris}.
\SW{Rebel} is often synonymous for \SW{Seth} 
\citep[][\LJnow{p91}]{Lei94}.  % RefOK

Note that the texts mentioning  \SW{Horus} are included in both 
Lists 1 and 2, because
the name \SW{Horus} appears in both mythical narratives
of Sects. \ref{Destr} and \ref{Conte}. 
However, the order of these \SW{Horus} texts is different
when rearranged in the order of increasing
$\Theta_{\mathrm{Algol}}$ or $\Theta_{\mathrm{Moon}}$.

\section{Discussion \label{discussion}}

We present one argument about CC in the end of each 
Sect. \ref{hourstars}-\ref{astrophys}.

\subsection{Measuring night-time with hour-stars
\label{hourstars}}

At the night-time in ancient Egypt, time was traditionally measured
from the positions of hour-stars.

\begin{figure*}[!t]\capstart
\centering
% kirjuri.tes
% cp kirjuri.eps final_porceddu_fig5.eps
\includegraphics[width=0.75\textwidth]{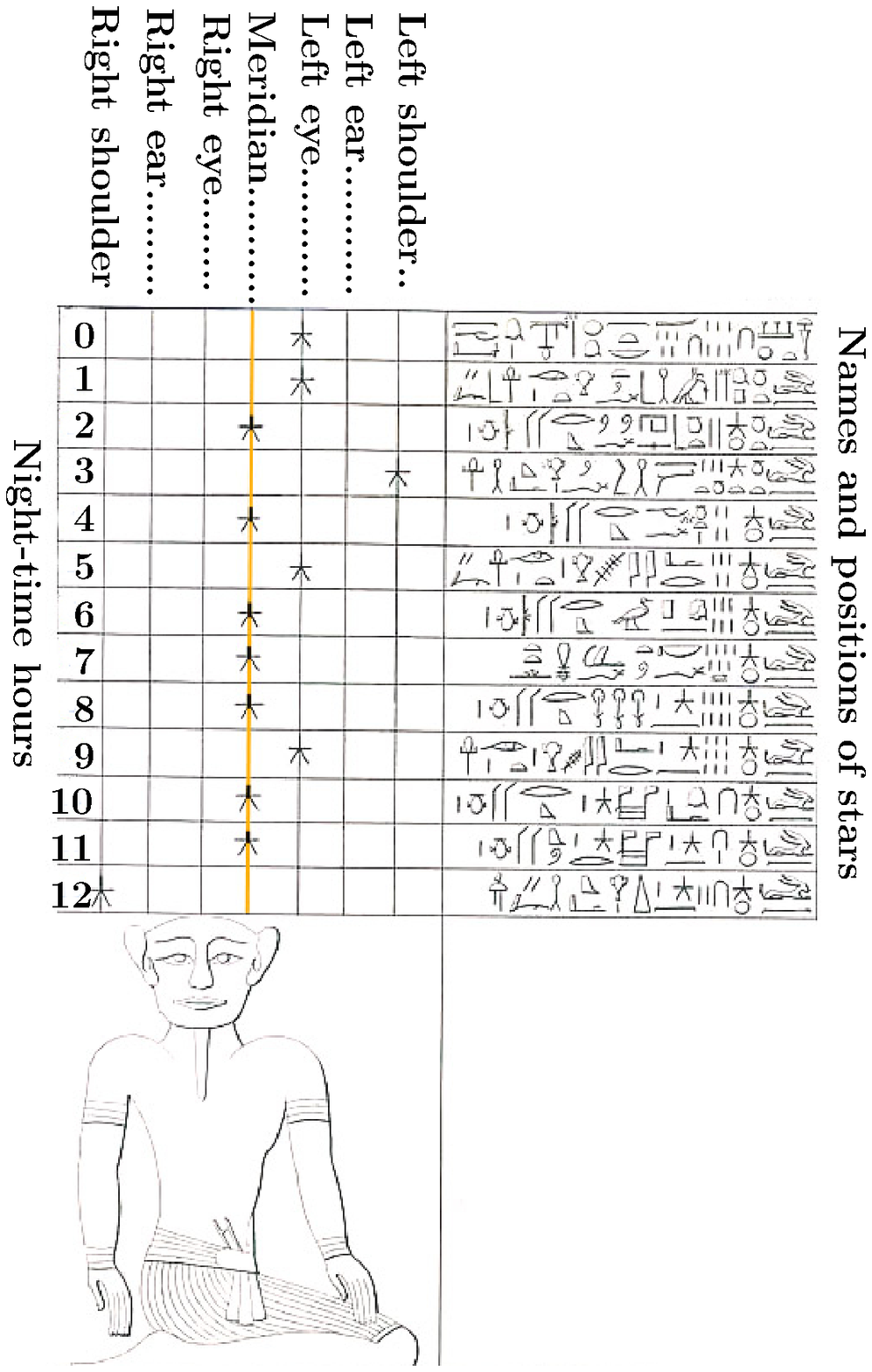}
\caption{{\bf Hour-watcher with a star chart
from the tomb of Ramesses VI, 12th century BC.}
``Meridian'' is ``opposite the heart'' mentioned
in Sect. \ref{hourstars}.
Reprinted under a  CC BY license @eng.wikipedia.org.}
\label{figfive} %fig5
\end{figure*}

The ancient Egyptian day was split into daytime and night-time, 
both with 12 hours. 
Time was counted using shadow clocks by day,
and star clocks or water clocks by night. 
The Egyptian 
``hour-watcher''\Egy 
was a specialized scribe whose job was to observe various 
hour-stars,
\textit{i.e.} clock stars whose positions began and ended 
the night hours (Figure \ref{figfive}). 
More specifically, a text previously known as 
``The Cosmology of Nut''
whose title 
was deciphered by \citet{Lie07}  to be 
``The Fundamentals of the Course of the Stars'' 
informs us that stars were traditionally observed 
when in positions of the 
``culmination''\Egy upon the first hour of the night 
(transit of the meridian between the eastern half of the sky 
and the western half of the sky), 
``heliacal setting''\Egy ~and 
``heliacal rising''\Egy \citep[][\LJnow{p56-65}]{Cla95}. % RefOk
Tabulated positions of 
hour-stars marked the closing of each night hour. 
One approximation 
for the length of an Egyptian night hour was 
40 minutes \citep[][\LJnow{p133}]{Lei95}. % RefOk

The Ramesside Star Clocks from the tombs of 
Ramses VI, Ramses VII and Ramses IX 
show an even more complex arrangement of 
hour-stars \citep[][\LJnow{p56-65}]{Cla95}.  % RefOk
Each table consists of thirteen rows of stars. 
The first row stands for the opening of the first night hour 
and the other twelve rows 
stand for the closing of each of the twelve night hours. 
In the rows, a star is positioned in respect 
to a sitting human figure. 
Possible positions are 
``upon the right shoulder''\Egy,
``upon the right ear''\Egy,
``upon the right eye''\Egy,
``opposite the heart''\Egy,
``upon the left eye''\Egy,
``upon the left ear''\Egy,
and
``upon the left shoulder''\Egy
~(Figure \ref{figfive}).
The system was originally interpreted as two hour-watchers 
opposite to each other on the roof of the temple precisely 
aligned to the line of the meridian. 
These timing observations would have been necessarily 
made towards south, because of the positions of the hour-stars.
The hour-watcher facing south might 
have utilized a plumb and a sighting device 
to determine when the given star 
is in exactly the right position and 
announce the closing of the hour. 
If the hour-watchers were positioned 2-3 meters away from each other, 
the slow rotation of the night sky 
would have provided a 10-15 minute difference between the marked 
positions \citep[][\LJnow{p33}]{Lei95}.
 According to 
\citet[][\LJnow{p165}]{Lul09}
it is more likely that the figure that had been interpreted as 
an hour-priest would rather be a divinity associated with time-keeping, 
leading to revised ideas regarding the direction of the observations: 
even some constellations of the northern half of the sky could 
have been used in this method of time-keeping. 
The references to the observational practices of 
the hour-watchers
are scarce and known mostly from late period sources 
such as the inscription on a statue depicting the astronomer Harkhebi 
and a sighting instrument with inscriptions mentioning 
an astronomer named Hor 
\citep[][\LJnow{p489-496}]{Cla95} and 
\citep[][\LJnow{p10-26}]{Pri10}. % RefOK
However, it is safe to say that such practices 
existed throughout Pharaonic history.
The observing conditions of the hour-watchers were rather ideal, 
with about 300 clear nights each year 
\citep[][\LJnow{pD7}]{Mik95}. % RefOK

Algol is the 60th brightest star in the sky \citep{Hof91}. % RefOK
At the latitude of Middle Egypt,
$\phi_{\mathrm{Earth}}=26.6^{\mathrm{o}}$,
the never setting circumpolar stars have declinations
$\delta_{\mathrm{Star}} > 90^{\mathrm{o}} - \phi_{\mathrm{Earth}}=63.4^{\mathrm{o}}$.
The declinations of stars that never rise above horizon are 
$\delta_{\mathrm{Star}} < \phi_{\mathrm{Earth}}-90^{\mathrm{o}}=-63.4^{\mathrm{o}}$.
Circumpolar stars are not ideal hour-stars, because they never
rise or set. Furthermore, their angular motion within a limited
area around the celestial pole is not ideal for measuring time.
It is not easy to obtain
accurate timing from the minor changes in their positions.
Stars below horizon can certainly not be used as hour-stars.
If the circumpolar stars and the stars below horizon are excluded,
Algol was the 56th brightest star at 
$\phi_{\mathrm{Earth}}=26.6^{\mathrm{o}}$ in 1224~B.C.
However, if a star culminates in the south 
below an altitude of $a=10^{\mathrm{o}}$,
its brightness decreases about one magnitude due to atmospheric
extinction and it can be observed only
for a short time even in ideal observing
conditions. Such a star is not a suitable hour-star.
A star that sets $10\aste$ (or less) below horizon in the north is neither
an ideal hour-star, because it rises and sets nearly at the same
location, and its brightness decreases close to the horizon due to extinction.
Using the limits $-53.4^{\mathrm{o}}<\delta_{\mathrm{Star}}<53.4^{\mathrm{o}}$,
makes Algol the 51st brightest star in the Middle Egyptian sky in 1224~B.C. 
These declination limits of ours are conservative,
because many other stars culminating in the north are useless
for time keeping, 
and extinction influences the comparison of the brightness levels
between Algol and the stars always remaining close to the horizon. 
This raises Algol much higher than
the 51st best in the list of suitable bright hour-stars
(FAQ \ref{nothourstar}).

Algol's equatorial coordinates were 
right ascension $\alpha_{\mathrm{Algol}}=1^{\mathrm{o}}$
and 
declination $\delta_{\mathrm{Algol}}=25^{\mathrm{o}}$ in 1224 B.C. 
which gives the following ecliptic coordinates,
longitude $\lambda_{\mathrm{Algol}}=11^{\mathrm{o}}$
and 
latitude $\beta_{\mathrm{Algol}}=22^{\mathrm{o}}$.
The ancient cultures used the latter coordinate system
based on the yearly motion of the Sun.
Algol was located very close to the vernal equinox 
and this ecliptic plane.
If the timing observations were made towards south,
then  the bright stars in 
the ecliptic plane were the most
suitable hour-star candidates
\citep[][\LJnow{p2-56}]{Cla95},
\citep[][\LJnow{p145-150}]{Wel01A}
and
\citep{Bok84}.
This location of Algol in the sky raises it very
high in the list of suitable bright hour-stars
(FAQ \ref{nothourstar}).
Furthermore,
Algol culminated at the altitude of $a=88^{\mathrm{o}}$,
and this made it an ideal star for measuring time.

One modern hour equals 15 degrees in the equatorial plane, 
and therefore the required minimum number of hour-stars
for covering the entire sky is at least 24. 
The earliest known star clock scheme, the 
so-called diagonal star clock, describes 36 decans 
\citep[][\LJnow{p63}]{Lei95}. 
For a ten-day week, 12 of the decans are tabulated 
marking the beginnings and ends of the hours. 
Because the sidereal day is about four minutes shorter 
than the solar day, these stars reach their positions 
four minutes earlier every consecutive night. 
After ten days the stars occupy their positions about 39 minutes earlier, 
so in the tabulation of decans for the next week 
each of the stars "works" one hour earlier. 
The concept of the hour was relative as its beginning 
was allowed to fluctuate about 35 minutes by the end of each week. 
The later Ramesside star clocks are comprised of a system of 
46 individual stars observed in positions 
such as ``upon the right shoulder'', ``upon the right ear'', 
etc. as in our Figure \ref{figfive} \citep[][\LJnow{p120}]{Lei95}. 
These positions were intended for better accuracy 
but the known tables were given for 15-day periods, 
where the timing of the hours fluctuated even 55  minutes. 
However, this model seems to have been outdated 
by the time it was used in the decoration of Ramesside tombs, 
so the later developments of the system 
remain unknown \citep[][\LJnow{p132}]{Lei95}.   

No-one can recognize a single hour-star
in the sky without comparing its position to 
the positions of other bright stars in its vicinity.
For the sake of consistency,
we introduce our own precise concept: ``hour-star pattern''.
Such a pattern contains one hour-star used by the scribes, 
and all bright stars that they used to identify this hour-star.
Using only two stars per one hour-star pattern 
would not have provided
any recognizable pattern.
Therefore, the number of stars per each hour-star pattern
must have been at least three, or probably more. 
This means that the Ancient Egyptians must
have observed at least $24 \times 3= 72$ bright stars, and in this case
the visual brightness of the 72th brightest 
star was about $2.^{\mathrm{m}}5$ in 1240~B.C.
Thus, it is certain that the 51st brightest star Algol
($2.^{\mathrm{m}}12$) with an ideal position in the night sky 
was included into some hour-star pattern (FAQ \ref{nothourstar}).
The names of hour-stars in the Ramesside star clocks include 
for example various different body parts and equipment of the Giant, 
the Bird and the Hippopotamus, suggesting that the stars were 
members of a known constellation 
(\textit{i.e.} an hour-star pattern).
According to \citet[][\LJnow{p157-194}]{Bel09},
a complete set of constellations 
formed the Egyptian celestial diagram, 
\textit{i.e.} every star belonged to some constellation.

Unlike Astronomy, Egyptology 
is not an exact science and 
few questions can be answered
with absolute certainty.
For our ten arguments I-X presented in
Sects. \ref{hourstars}-\ref{astrophys} it is not important 
if Algol was an actual hour-star
or only a member of 
some hour-star pattern or related constellation (FAQ \ref{nothourstar}).
Algol has not yet been unambiguously identified in 
any hour-star lists 
because only the names 
of Sirius, Orion and the Plough 
have reached a widespread consensus among egyptologists.
Algol ($\beta$ Persei) is the second brightest star 
in the modern constellation of Perseus.
Egyptologists 
have presented their own differing 
identifications of the represented stars with actual stars so it 
is difficult to say which decan or group would have included Algol
\citep[][\LJnow{p161-162}]{Bel09},
\citep{Bok84}
and
\citep{Con03}. 
\citet[{p157-158}]{Lul09} claim to have uncovered 
nearly three quarters of the Egyptian firmament by 
deciphering the star names of aforementioned tomb of Senenmut, 
the clepsydra (water clock) of Amenhotep III and the circular zodiac 
of the temple of Hathor at Dendera, which is from Late Period and 
already incorporates Mesopotamian and Greek influences.
According to them,
the ancient Egyptian constellation of the Bird 
includes the modern constellations Triangulum and Perseus 
but there is no precise identification of the individual stars of the Bird.
\citet{Bok84} % RefOK
suggested that the correct reading for the decan Khentu, 
a group of three stars, is 
the ``snorting one''\Egy. 
The decan is later depicted as a red-haired warrior 
with fierce attributes reminiscent of Perseus in Greek mythology. 
The decan is also known as 
``the lower Khentu''\Egy, 
and mentioned in the decan lists of the Astronomical ceiling 
of the tomb of Senenmut (ca. 1473 B.C.), 
tomb of Seti I (1313-1292 B.C.) and 
the Osireion, a temple in Abydos 
dated to the time of Seti I \citep[][\LJnow{p23-26}]{Neu60}. % RefOK

The number seven seems to carry plenty 
of mythological connotations for the Egyptians, 
such as the seven failed attempts of \SW{Seth} to lift the foreleg 
into the heavens \citep[][\LJnow{p28}]{Lei94}.  % RefOK
It is reminiscent of the various names of the Pleiades, 
a distinct open cluster of bright stars located near 
Perseus (\textit{i.e.} Algol), 
known for example as Seven Sisters, Starry Seven 
and Seven Dovelets \citep[][\LJnow{p391-403}]{All99}. % RefOK
Pleiades may have been connected to the
\SW{Followers} or the \SW{Ennead} which were both
connected to $P_{\mathrm{Algol}}$ 
in CC \citep[][\LJnow{p14-15}]{Jet15}. % RefOK

A list of decan deities in papyrus Carlsberg I mentions 
certain stars that cause ``sickness''\Egy in fish and birds, 
while CC speaks of "a star with bitterness in its face". 
According to \citet[][\LJnow{p307}]{Lei94}, % RefOK
``bitterness''\Egy is the name of a sickness 
that plagued the Egyptians. 
Decans in the inscription 406 of 
the temple of Esna are portents of death to the 
"rebel" \citep[][\LJnow{p48}]{Lie00}.  % RefOK
Thus, even without knowing which stars exactly 
they referred to in these passages, 
we may conclude that the ancient Egyptians strongly believed 
in stars influencing the lives of men.

Argument I:
For thousands of years, the ``hour-watchers'' practiced the tradition
of timekeeping by observing hour-stars.
If Algol was not an hour-star, it certainly 
belonged to some 
hour-star pattern or related constellation.

\subsection{Crucial timing of nightly rituals \label{rituals}}

Proper timing was considered crucial
for the efficacy of nightly religious rituals.

Astronomy is often considered to be one of the oldest sciences
practiced by mankind despite ancient star 
observing being carried out for the benefit of 
religious practices.
Babylonians were probably the first people 
to make systematic notes of the Moon and the planets and also
to perform calculations of their celestial motions. 
As opposed to this, ancient Egyptian records 
that span three millennia 
are almost obsolete in any quantitative approach 
despite the culture's special attention to the Sun, the Moon, 
the planets and the stars as divine entities 
\citep[][\LJnow{p71-72}]{Neu51}. % RefOK
The phases of the Moon and 
the heliacal rising of Sirius played a part 
in determining the date and time of New Year and 
several other important festivals.

In ancient Egypt, the scribal professions were the most valued ones, 
as the entire functioning of the highly developed culture
and state with its complex bureaucracy 
was based on written communication \citep[][\LJnow{p24-38}]{Sha12}. % RefOK 
Many of the professional scribes had several titles 
emphasizing their specialized knowledge \citep[][\LJnow{p18-24}]{Cla89}, % RefOK
such as 
 ``physician''\Egy, 
``healer''\Egy, 
``hour-watcher''
 (astronomer who observed stars for timekeeping purposes) 
 and ``mathematician''\Egy. 

Beside the religious background, 
the scribes had plenty of social, 
political and personal motivation 
to perform their job with utmost expertise. 
Many scribes received the title of ``king's favourite'' 
during their lifetime and such persons 
were 
among the most high ranking 
members of the Egyptian society \citep[][\LJnow{p195}]{Cla89}. % RefOK

What would have been the ancient Egyptian scribes' interest
 in the behaviour of a variable star? 
Knowing the period and the phase of the Moon 
was important for regulating the religious festivities 
but the scribes would also have paid attention 
to any 
unexpected
changes in the observed 
hour-star patterns.
The hour-watchers' activity 
required the mapping and 
measuring of the heliacal rising and setting,
as well as the meridian transit of stars 
\citep[][\LJnow{p55}]{Mag13}. % RefOK
The average width for a suitable ancient Egyptian 
hour-star pattern
would have been about one hour in modern right ascension, \textit{i.e.} 15 degrees.
Thus, during every year, Algol's hour-star pattern 
determined the beginning of the night during half a month 
($\sim 15\aste/360\aste$),
the epoch of midnight during another half a month,
and the end of the night during yet another half a month.
During thousands of years of timing observations, the
discovery of Algol's variability would have been most probable
during 
these particular 
time intervals of every year (FAQ \ref{nothourstar}).

It is essential to note that the professional class of scribes 
was responsible for {\it both} astronomical observations 
and religious traditions. 
To be involved in the science of Astronomy (\textit{e.g.} hour-watching)
was also to be involved in the priesthood  (\textit{e.g.} nightly rituals) 
\citep[][\LJnow{p11-16}]{Sha12}. % RefOK
To properly observe the ritual cycle and 
recite the magical words at the exactly right time 
was of foremost importance to the Egyptian priests, 
since it was a matter of life and death 
\citep[][\LJnow{p64-68}]{Ass01}. % RefOK
The hour-stars were used for this exact timing that
was crucial for keeping cosmic order 
\citep[][\LJnow{p195}]{Cla89}
and
\citep[][\LJnow{p2}]{Mag13}. % RefOK
During the night, the Sun 
was considered to sail across the underworld 
where the prayers or incantations of the priests 
opened the gates of the underworld and 
appeased the terrible guardians of the gates 
\citep[][\LJnow{p37-58}]{Wie07}.  % RefOK
If everything went absolutely right, 
the Sun was reborn on the 12:th hour of the night. 
Any failure by the priests in observing the nightly rituals 
would mean running a 
risk of the Sun not rising the next morning 
\citep[][\LJnow{p68-73}]{Ass01}. % RefOK

The consequential purpose of astronomical observations
was ``religious astronomy'' \citep[][\LJnow{p188}]{Lie00}.
Any unpredictable change in the 
hour-star patterns observed 
by the scribes would have been a shock. 
The ritual activity of the ancient Egyptian priesthood 
functioned for the very purpose of 
maintaining the known universe in a stable condition in order 
not to plunge into chaos. 
They needed to use all their resources 
to keep ``Maat''\Egy,
the cosmic order \citep[][\LJnow{p2}]{Mag13}. % RefOK
If they found out that a star is variable in brightness, 
observing its cycle would most certainly 
have gained their extra attention and intellectual effort.

Argument II:
Proper timing of the nightly religious rituals
relied on the fixed hour-star patterns.

\subsection{Constellation change }

Any unpredictable
change in the fixed and known 
hour-star patterns
would have been alarming.
A naked eye observer witnesses
a radical 
hour-star pattern change
during Algol's eclipse.

The naked human eye can detect brightness differences of
0.1 magnitudes in ideal observing conditions.
Hence, naked eye eclipse detection is theoretically 
possible for 7 hours when Algol 
is more than 0.1 magnitudes dimmer than its brightest suitable comparison
star $\gamma$ Andromedae 
\citep[see][their Figure 5a]{Jet13}. % RefOK 
For 3 hours, Algol is also dimmer 
than its other five suitable comparison stars
$\zeta$ Persei, 
$\eta$ Persei, 
$\gamma$ Persei, 
$\delta$ Persei and 
$\beta$ Trianguli.
The detection of
Algol's eclipse is easy 
during this 3 hour time interval.

With $P_{\mathrm{Algol}}=2.850=57/20$ days,
these opportunities for easy detection follow
the sequence of ``3+3+13=19'' days \citep[][\LJnow{p20-21}]{Jet15}. % RefOK
A plausible hypothesis is that the ancient Egyptians first 
discovered the variability of Algol, like Montanari did in 1669, 
when they were observing Algol's 
hour-star pattern.
During primary eclipses, Algol lost its brightness gradually 
for five hours until it was outshined  by
its six dimmer nearby comparison stars.
The dimming was followed by a brightening 
that lasted for another five hours.
This entire 10 hour eclipse event 
could be observed during a single night, 
but that was rare event, because it occurred only every 19th night.
Algol's 
hour-star pattern change was so noticeable
that the priests on duty as hour-watchers
could hardly have missed this event.
For the hour-watchers, 
it would have been useful 
to communicate among themselves the knowledge 
about the strange behavior of this 
hour-star pattern and 
there is a good possibility, 
considering the diligent scribe mentality of Egyptian officials, 
that they would have made written notes about the times of the eclipses.

The modern constellation of Perseus is one of the easiest to perceive. 
It occupied a prominent position high in the ancient Egyptian night sky,
because the maximum altitude of Algol was 88 degrees.
Algol is the brightest member of another ancient constellation 
of four stars which was already recognized by authors like 
Vitruvius \citep[][\LJnow{p266-267}]{Vit60}    % RefOK 
and Ptolemy \citep[][\LJnow{p31}]{Pto15}. % RefOK
The other three stars are 
$\pi$ Per ($4.^{\mathrm{m}}7$), 
$\omega$ Per ($4.^{\mathrm{m}}6$)
and 
$\rho$ Per ($3.^{\rm m}4-4.^{\mathrm{m}}0$). 
During the primary eclipses, 
the brightness of Algol falls from $2.^{\rm m}1$ to $3.^{\rm m}4$, 
\textit{i.e.} these other three stars never appear to be brighter than Algol. 
The shape of this constellation resembles a diamond 
and it was therefore called the Head of Gorgon 
or the Head of Medusa \citep[][\LJnow{p332}]{All99} % RefOK
in the Hellenistic culture. 
The angular separation between Algol 
and the other three stars is less than two degrees, 
and this constellation is therefore ideal for detecting variability
because
atmospheric extinction 
does not mislead brightness comparisons even at low altitudes
close to the horizon. 
At its brightest, Algol visually dominates this diamond shaped constellation,
because it is clearly much brighter than the other three stars.
Hence, a naked eye observer can easily notice the
significant constellation pattern change during Algol's eclipse.  
\citet{Wil96}
suggested that from this may have arisen the myth of the 
Medusa losing its head. % RefOK

In principle, 
other variable stars besides Algol, 
like the disappearing and reappearing Mira, also called Omicron Ceti, 
might also have been discovered by the ancient Egyptians. 
However, the eleven month period of Mira is so
long that it can not be rediscovered with statistical methods
in CC \citep[][p9, see Criterion $C_2$]{Jet13}. % RefOK 

Argument III: A naked eye can easily
discover the significant hour-star pattern
change caused by Algol's eclipse.

% jha/Period19.py produces Period19.eps
% cp Period19.eps final_porceddu_fig6.eps
%%BoundingBox: 90 330 558 504
\begin{figure*}[!t]\capstart
%\centering
\includegraphics[width=\textwidth]{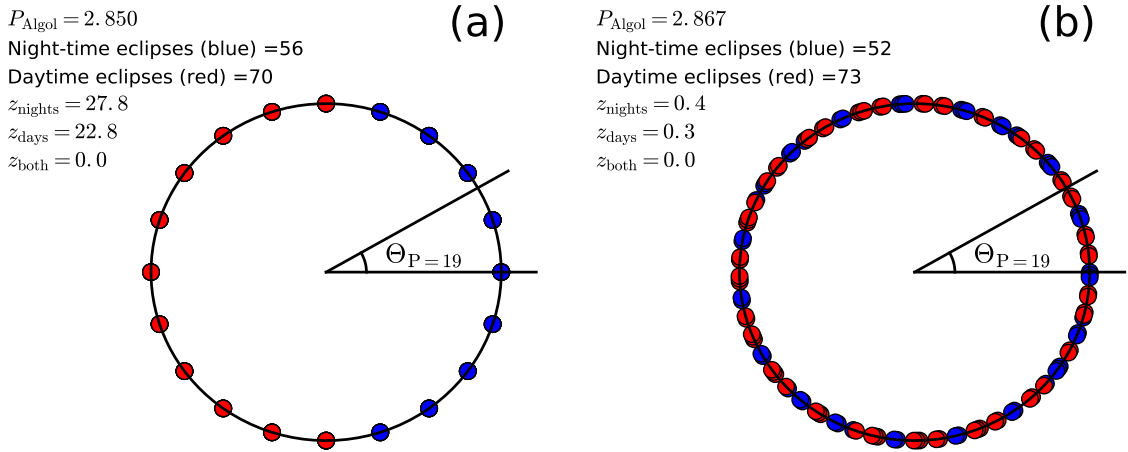}
\caption{
{\bf Mid-epochs of Algol's primary eclipses.}
{\bf (a)} 
The yearly mid-epochs of 56
night-time eclipses occurring $\pm 5$ hours at
both sides of midnight with the period $P_{\mathrm{Algol}}=2.850$ days
(blue circles).  The red circles denote the 70 daytime eclipses.
Note that only 20 circles are visible,
because the 56 red or 70 blue circles overlap.  
The eclipse phase angles $\Theta_{\mathrm{P=19}}$ are computed
for the period $P=19$ days.
The values of the Rayleigh test 
statistic for these eclipse epoch time points
with $P=19$ days
\citep[][$z$ in Eq. 7]{Jet15} are given for 
the night-time eclipses $(z_{\mathrm{nights}})$,
the daytime eclipses $(z_{\mathrm{days}})$
and
all eclipses $(z_{\mathrm{both}})$.
{\bf (b)} The same as in (a), except that the current 
period value, $P_{\mathrm{Algol}}=2.867$ days, is used.
}
\label{figsix} %fig6
\end{figure*}

\subsection{``3+3+13=19'' and ``19+19+19=57''
days eclipse rules  \label{eclipserules}}

Algol's night-time eclipses followed the regular
3+3+13=19 and 19+19+19=57 days cycles
with $P_{\mathrm{Algol}}=2.850=57/20$ days.
\citep{Jet15}. % RefOK

If an eclipse first occurred in the end of the night, 
then after three days the next eclipse occurred close to midnight.
After another three days the next eclipse
occurred in the beginning of the night.
After these three eclipses, it took
thirteen days, before the next eclipse was observed,
\textit{i.e.} the first eclipse in the end of the night was repeated 
again \citep{Jet15}. % RefOK
This is the ``3+3+13=19'' days rule, \textit{i.e.} 
a sequence of three night-time eclipses was repeated every 19 days. 
The midnight eclipses,
when both the dimming and the brightening 
of Algol could be observed during the same night,
occurred at 19 days intervals.
Coincidence or not, 
the CC prognosis of the first day of every month is GGG 
and the prognosis of the 20th day of every month is SSS.
These two particular days of 
{\it every} month
are separated by the ``3+3+13=19'' days eclipse cycle.

With the period of $P_{\mathrm{Algol}}=2.850=57/20$,
eclipses separated by 57 days would have occurred 
at exactly the same time of the night. 
This is the ``19+19+19=57'' days rule, \textit{i.e.} 
this entire 57 days eclipse sequence kept on 
repeating itself over and over again.
For this reason,
there are only 20 small blue or red circles (night-time or daytime eclipses)
in Figure \ref{figsix}a, 
although this figure shows all 126 eclipses occurring
during 360 days.
If the hour-watchers began to make notes of the eclipse epochs,
the ``3+3+13=19'' days eclipse cycle would have been discovered first, 
then later the ``19+19+19=57'' days eclipse cycle. 
The first regularity is easier to discover, 
because all Algol's night-time eclipses 
follow this ``3+3+13=19'' cycle with $P_{\mathrm{Algol}}=2.850=57/20$ days. 
The second cycle of ``19+19+19=57'' days is more difficult to discover, 
because it requires the measurement of 
the nightly shifts of these night-time eclipses within the ``3+3+13=19'' cycle.
Although the ancient Egyptians had no way of knowing 
that Algol's eclipses also happened during daytime,
their long-term eclipse records may
have eventually led to the discovery of this 57/20 days ratio.

Simulations have shown
that modern period analysis would rediscover the 2.850 days period in CC,
if the scribes had recorded nearly 
all observed night-time eclipses \citep{Jet13}. % RefOK
The fact that the 19 days period is also present 
in these data \citep{Jet13} % RefOK
confirms that {\it only} 
the night-time eclipses were used in the construction of CC.
This result can be deduced from Figure \ref{figsix}a.
It is highly unlikely that the ancient Egyptians knew 
that Algol's eclipses also occurred at daytime,
because these events could not be observed.
If an eclipse occurred at the daytime, 
then no eclipse was observed on the previous night or the next night. 
However, this absence of night-time eclipses could be observed.
This absence was observed in 16 nights of every 19 days cycle.
Our Figure \ref{figsix}a  shows that there are two plausible alternatives
that exclude each other.
Either the {\it presence} (Figure \ref{figsix}a: $z_{\mathrm{nights}}=27.8$)
or {\it absence} (Figure \ref{figsix}a: $z_{\mathrm{days}}=22.8$)
of night-time eclipses was used as a criterion for selecting 
the lucky prognoses connected to Algol's variability.
In other words,
if the ancient Egyptians had used {\it both} 
the night-time and the daytime eclipses in the construction of CC,
this would have erased the 19 days signal completely away from CC
(Figure \ref{figsix}a: $z_{\mathrm{both}}=0.0$).
The alternative that the absence of night-time eclipses 
was used in assigning lucky prognoses is more probable,
because the lucky prognoses mentioning \SW{Horus}, \SW{Sakhmet}
and \SW{Wedjat} concentrate on the bright phases of Algol
(Figure \ref{figone}). This alternative is later discussed 
in greater detail in Sect. \ref{Reasoning} (principle I). 

The scribes seem to have applied a very accurate value for the synodic
period of the Moon, 29.6 days, to correctly predict the lunar phases
\citep{Jet13}.
If they intended a correct prediction of Algol's eclipses 
they would have also needed an accurate period value.
Our Figure \ref{figsix}b shows how the ``3+3+16=19'' days 
cycle breaks down with the currently observed 2.867 days period of Algol.
Modern period analysis detects the 19 days
and the 2.850 days periods in CC,
but it does not detect 
the current 2.867 days period of Algol \citep{Jet13}.
Our Figure \ref{figsix}b confirms that if the period of Algol in those days
had been the same as the current period, 2.867 days,
there would be no signs of the 19 days period in CC
($z_{\mathrm{nights}}=0.4$, $z_{\mathrm{days}}=0.3$ and $z_{\mathrm{both}}=0.0$).

Argument IV: The scribes could have discovered Algol's 
$2.850=57/20$ days period
from long-term observations of 
the regular 19 and 57 days eclipse cycles.

\subsection{Ancient Babylonian and 
Chinese lunar eclipse prediction}

The ancient Egyptian discovery of $2.850=57/20$ days ratio in 
Algol's eclipses
would resemble
the ancient Babylonian and Chinese discoveries of the lunar eclipse cycle.
Here, we discuss {\it how} these two analogous
ancient astronomical cycle discoveries were made,
and {\it how} the ancient Egyptians may have utilized 
a similar approach to observe Algol's variability.

Firstly,
the process that may have lead to the ancient Egyptian 
discovery of the periodicity of Algol's eclipses 
could have been analogous  to the process 
which led to the ancient Babylonian 
detection of the cycle in the lunar eclipses. 
The unexpected occurrence of solar and lunar eclipses 
must have deeply impressed ancient civilizations. 
The motivation for prognostications was obvious. 
There is a description of how the Babylonians 
determined the Saros cycle, 18 years and 11.33 days, 
in the occurrence of lunar eclipses \citep[][\LJnow{p57-62}]{Pan89}. % RefOK
Detailed long-term records were kept of these events. 
Somewhere between 750 and 650 B.C. there 
was a more or less complete record of the observed eclipses 
and by fairly basic analysis it 
was found out that within 223 lunar months there 
were 38 eclipse possibilities. 
Also these phenomena always occurred 
in a series of four, five or six consecutive lunar eclipses. 
A theory of the past and future lunar eclipses, 
which was based on the table called ``Saros-Canon'', 
emerged sometimes after 280 B.C. 
\citet[][\LJnow{p181}]{Bra05} % RefOK
have argued that the Babylonians 
may even have invented sophisticated methods 
to determine the anomalies (between 6 and 9 hours) of the Saros cycle.
Ancient Egyptians were also fluent 
in arithmetic and geometric calculations 
and devised plenty of tabulations 
to aid in practical life 
\citep[][\LJnow{p24-42}]{Cla99}
and
\citep[][\LJnow{p57-60}]{Ros04}. % RefOK 
Papyrus Carlsberg 9 reveals that they used 
long-term observations
to determine the correct period of the Moon
\citep[][\LJnow{p23-28}]{Cla95}. % RefOK 
It has also been shown
that the $P_{\mathrm{Moon}}$ value 
in CC was so accurate it must have been determined
from observations made over more than one year 
\citep[][\LJnow{p7}]{Jet13}. % RefOK 
These two cases \citep{Jet13,Cla95} % RefOK
confirm that the ancient Egyptians were capable of using long-term
observations to determine the periods of celestial objects.
By the Late Period, the Egyptians 
were also familiar with the use of the Saros cycle of 223 synodic 
months to predict lunar and solar eclipses.
Ptolemy's commentators referred to these periods as 
having been 
used already by ``the ancients'' \citep[][\LJnow{p88}]{Ste00}. % RefOK

Secondly, at about 400 B.C.
the Chinese 
were using an astronomical calendar to predict 
the dates of lunar eclipses \citep[][\LJnow{p175-178}]{Ste00}. % RefOK
This calendar made use of a cycle of 135 months 
which includes 23 lunar eclipse possibilities. 
They approximated that eclipse possibilities 
occur periodically every 5 and 20/23 months and 
counted them arithmetically.
The ancient Egyptians could have used an analogous
approach to Algol's eclipses.

The alternating period changes of Algol are so small 
\citep{Bie73} % RefOK
that they do not mislead long-term period determination
based on naked eye observations.
However, the synodic period of the Moon varies between 29.3 and 29.8
days in a year \citep{Ste91}. % RefOK
Thus, Algol's eclipses are far easier to predict 
than lunar eclipses, because Algol's period is constant, 
while that of the Moon is not.
The 2.850=57/20 days ratio means that 20 
eclipses of Algol
occur during 
{\it every} 57 days cycle (Figure \ref{figsix}a). 
If the scribes performed
long-term observations of these cycles, 
these observations
would have eventually revealed that 
{\it all} nine Algol's night-time eclipses could {\it always} be observed
within {\it every} 57 days cycle.

There are three alternative methods of {\it how} the period
of 2.850 days ended up into CC.
We may never find out which one of these three methods was used.
However, in every alternative the scribes would have discovered the eclipses 
of Algol, and recorded the period of these events.

\begin{itemize}

\item[ 1st] ~method: The scribes might have calculated the numerical value 
$P_{\mathrm{Algol}}=2.850$ days from long--term observations.

\item[ 2nd] ~method:
The scribes may not have had any reason for
even trying to calculate the actual period.
If they noticed that Algol's three consecutive night-time eclipses 
followed the 19 days cycle, while nine consecutive night-time eclipses 
occurred in every 19+19+19=57 days cycle,
these cycles would have appeared extremely stable over very
long periods of time.
The minor nightly shifts in the epochs of the 
eclipses within individual 19 days cycles
cancelled out within every 57 days cycle.
The scribes may not even have noticed these minor nightly shifts 
within individual 19 days cycles, 
because the eclipses at the end of the night, 
close to the midnight or at the beginning of the night always returned
back exactly to the same moments of night after every 57 days cycle. 
If the real period of Algol was 57/20=2.850 days,
these 19 and 57 days cycles would have enabled the scribes
to foretell what would be observed in the sky.
While the scribes may have 
recorded into CC {\it only} the 19 or 57 days cycles
in the night-time eclipses of Algol,
modern period analysis confirms that these rules worked 
perfectly {\it only if}
the real period of Algol was 57/20=2.850 days in those days.
Hence, the scribes have, although only perhaps unknowingly, 
also recorded the 57/20=2.850 days
period into CC.

\item[3rd] ~method: The scribes only
used the observed epochs of night-time eclipses 
as such in the construction of CC.
This could be achieved even without ever 
solving a numerical estimate of $P_{\mathrm{Algol}}=2.850$ days,
or without ever discovering the 19 and 57 days cycles.
Also in this case,
the scribes have unknowingly
recorded the 57/20=2.850 days period into CC.

\end{itemize}

Argument V:
The ancient Egyptian scribes may have calculated
the $57/20=2.850$ days period of Algol from long-term
 observations (1st method).
 They may not have calculated this 2.850 days period,
because the 19 days and 57 days cycles already 
perfectly predicted {\it all} night-time eclipses of Algol (2nd method), 
or they may have just recorded the observed night-time eclipses into CC
(3rd method).

\subsection{Indirect references to protect cosmic order}

After the scribes had discovered Algol's variability, 
it would have been attributed religious significance 
and described accordingly.

Lack of direct references to Algol's variability in ancient Egyptian 
records raises questions. 
We may draw parallels to solar eclipses, 
which were experienced by Egyptians during the New Kingdom 
and could not have passed unnoticed. 
Yet, these events are not mentioned directly in 
written records \citep{Smi12}. % RefOK
The first plainly written Egyptian eclipse records 
are found in the demotic papyri Berlin 13147 + 13146 
which date to the first century B.C. \citep[][\LJnow{p85-91}]{Ste00}. % RefOK
\citet{Smi12} % RefOK
interpreted references 
to solar eclipse events in New Kingdom texts and 
concluded that astronomical events 
were described indirectly by using religious terminology and 
the reference might even have been made deliberately obscure. 
In Late Period demotic writings eclipses were mentioned 
and described as fearsome and unlucky portents 
\citep[][\LJnow{p102}]{Lie99}. % RefOK

In general, ancient Egyptian scribes seem 
to have avoided direct references to celestial events,
because writing was considered in itself to have a magical power 
that allows the scribe to communicate with the gods. 
This could be the reason why they avoided direct 
references to the celestial events,
\textit{i.e.} the observed actions of divine deities in the sky.
The scribes would have avoided making direct textual references 
to the uncanny behaviour
of Algol in order to preserve cosmic order. 
Divine names that may have been related to Algol are various, 
depending on the context. 
In particular, the names \SW{Horus}, \SW{Wedjat} and \SW{Sakhmet} 
had similar phase angle $\Theta_{\mathrm{Algol}}$ distributions in 
Figures \ref{figone} and \ref{figtwo}.
We  conclude that references to astronomical events
are indirect but undoubtedly present throughout the whole CC text, 
not unusual for Egyptian mythological texts in general
\citep[][\LJnow{p285-286}]{Lei94}. % RefOK 

Argument VI: To avoid violating cosmic order,
the scribes would have referred to Algol's changes only indirectly.

\subsection{Two legends  \label{Legends}}

Lists 1 and 2 are {given in Appendix} C and D, respectively.
These lists are full of extracts from two well known legends: 
``the Destruction of mankind'' (hereafter LE1)
and ``the Contendings of Horus and Seth'' (hereafter LE2).
CC texts seem to refer to many lesser known legends as well,
 but most of the extracts on these lists, 
chosen by the occurrence of the selected SW, 
are clearly connected to these two well known legends 
(for example "peace between Horus and Seth", "pacify the Wedjat"). 
Those without a clear connection to either of these two such 
as "Horus hears your words in front of every god" or "Horus 
is proceeding while Deshret sees his image" could fit into other narratives. 
However, these are of generic or fragmentary nature and the associations 
would be too speculative to help to understand the significant 
periodicity in the prognoses. It is not our ambition here 
to explain the prognosis of all individual dates 
but to find the logic behind the periodicity, 
caused by a larger group of connected dates and prognoses.

\subsubsection{Destruction of Mankind (LE1): \label{Destr}} 

In this legend, Re sends the Eye of \SW{Horus} (\textit{i.e.} \SW{Wedjat})
to punish the rebellious mankind.

The legend of the ``Destruction of Mankind'' 
is a mythological narrative that figures 
repetitively in CC.
It concerns the Eye of \SW{Horus}, 
also called \SW{Wedjat} or 
``The Raging One''\Egy
\citep[][\LJnow{p361}]{Lei03}, % RefOK
fighting against the rebels who oppose the Sun god \SW{Re}. 
In the beginning of this legend, 
Re sends auxiliary gods in the form of fishes 
to overhear the plots of the rebellious mankind 
\citep{Gui03}. % RefOK
The impudence of 
mankind causes \SW{Re} to send the Eye of \SW{Horus} to kill all the rebels. 
As the Eye takes the form of the lion-goddess \SW{Sakhmet}, 
the destruction of the entire mankind is imminent. 
The gods deceive the goddess \SW{Sakhmet} by colouring beer mash 
with hematite to make it look like human blood. 
\SW{Sakhmet} drinks the beer mash, is pacified and 
mankind is saved \citep[][\LJnow{p197-199}]{Lic76}. % RefOK

\subsubsection{Contendings of Horus and Seth (LE2): \label{Conte}}

In this legend, Horus and Seth contend for the kingship of Egypt.

Another explicitly quoted legend in CC is 
the ``Contendings of Horus and Seth''.
 After being murdered, the divine ruler \SW{Osiris} 
is in the underworld and the contenders for his office 
are his son \SW{Horus} and his brother \SW{Seth}, 
who was responsible for the death of \SW{Osiris}. 
The dispute is decided by the council of nine gods 
called the \SW{Ennead}, ruled by the Sun god \SW{Re}. 
Various contests are ordered for \SW{Horus} and \SW{Seth} 
to determine who 
is the able and rightful ruler of Egypt,
\SW{Horus} being described as physically weak but clever, 
\SW{Seth} stronger but with limited intelligence. 
\SW{Seth} is defeated and two parallel judgements conclude the myth. 
The first verdict of the gods is a division of the kingdom 
between the two, 
but the second verdict is gods 
giving \SW{Horus} the entire inheritance of his 
father \SW{Osiris} \citep[][\LJnow{p294-295}]{Red01}. % RefOK 
Due to parallel judgements,
the legend ends either to the crowning of both \SW{Horus} and \SW{Seth},
or only of \SW{Horus}.

Argument VII: Even a quick glance on
List 1 ($\Theta_{\mathrm{Algol}}$ order)
and
List 2 ($\Theta_{\mathrm{Moon}}$ order)
reveals that numerous CC texts are excerpts
from the LE1 and LE2 legends.

\subsection{Astronomical beliefs behind Lucky and Unlucky Days 
\label{Reasoning}}

The scribes used the LE1 and LE2 legends to describe the phases 
of Algol and the Moon. 
Here, we show how the events of LE1 and LE2 appear in Lists 1 or 2.
This reveals the three principles that they 
possibly used to describe celestial variability as 
activity of deities.

The periods  $P_{\mathrm{Algol}}$ and $P_{\mathrm{Moon}}$ were discovered
from over 500 time points of lucky prognoses \citep{Jet13}. % RefOK
These large samples were used to determine the
zero epochs $t_0$ of ephemerides Eq. \ref{Aephe} and \ref{Mephe}.
The phase angles of lucky prognoses concentrated at these epochs
$\Theta_{\mathrm{Algol}}=0\aste$ at Aa in Figure \ref{figone}
and $\Theta_{\mathrm{Moon}}=0\aste$ at Ma in Figure \ref{figthree}.
This means that these two particular phase angles 
were considered to be the luckiest in
the cycle of Algol and the Moon, respectively.

When the CC passages are read in temporal order from one day 
to the next, the general sequence of events appears disorganized.
However, the contents of the passages begin to make sense when
these passages are rearranged and read in the order of increasing 
$\Theta_{\mathrm{Algol}}$ and $\Theta_{\mathrm{Moon}}$ 
(Lists 1 and 2).
This result could have been accomplished even without
solving the zero epochs $t_0$ of Eqs. \ref{Aephe} and \ref{Mephe},
because the same stories are repeated
in the same phase angle order of Algol and the Moon.
In other words, the $\Theta_{\mathrm{Algol}}$ and $\Theta_{\mathrm{Moon}}$ 
order of these CC passages is unambiguous.

The first principle in assigning  Lucky and Unlucky Days seems to have been 
\begin{itemize}

\item[] principle I:
 {\it The middle of the
bright phases of Algol and the Moon is lucky for mankind.}

\end{itemize}

For $P_{\mathrm{Algol}}$,
point Aa denotes the luckiest phase angle 
$\Theta_{\mathrm{Algol}}=0\aste$ in Figure \ref{figone}.
Out of all 28 SWs studied by \citet{Jet15}, % RefOK
the $n=14$ lucky time points $g(D,M)$ of \SW{Horus} have the strongest 
amplifying impact on the $P_{\mathrm{Algol}}$ signal 
(Figure \ref{figone}: closed squares).
Twelve of these fourteen values, having $-90\aste <P_{\mathrm{Algol}}<+90 \aste$,
amplify this signal.
The closer the phase angle $\Theta_{\mathrm{Algol}}$ of any  
$g(D,M)$ is to $\Theta_{\mathrm{Algol}}=0\aste$,
the more this time point amplifies the $P_{\mathrm{Algol}}$ signal.
Here are short excerpts from the five lucky \SW{Horus}
passages closest to $\Theta_{\mathrm{Algol}}=0\aste$
\setlist[itemize]{topsep=5pt,leftmargin=45pt}
\setlist[enumerate]{topsep=5pt, leftmargin= 45pt}
\begin{itemize}
\item[$g(14,2)$] $\equiv \Theta_{\mathrm{Algol}}=6\aste$,
{\it ``the majesty of Horus receiving the white crown'' } 

\item[$g(19,12)$] $\equiv \Theta_{\mathrm{Algol}}=13\aste$,
{\it ``this eye of Horus has come, is complete, is uninjured'' } 

\item[$g(27,1)$] $\equiv \Theta_{\mathrm{Algol}}=19\aste$,
{\it ``Peace between Horus and Seth''}

\item[$g(24,3)$] $\equiv \Theta_{\mathrm{Algol}}=19\aste$,
{\it ``Onnophris' happiness in giving his throne to his son Horus'' } 

\item[$g(1,7)$] $\equiv \Theta_{\mathrm{Algol}}=32\aste$,
{\it `` a feast of entering into heaven {\rm (\textit{i.e.} appearing into the sky)}.
The two banks of Horus rejoice'' } 

\end{itemize}

\noindent
These texts suggest that the lucky prognoses of \SW{Horus}
were connected to the bright phases of Algol \citep{Jet15}. % RefOK

For $P_{\mathrm{Moon}}$,
the luckiest phase angle $\Theta_{\mathrm{Moon}}=0\aste$ coincides
with point Ma in Figure \ref{figthree}.
\citet[][\LJnow{p285-286,474}]{Lei94} % RefOK
 had already argued that the New Moon occurred between 
the unlucky time points on the first of the following two consecutive days
\begin{itemize}

\item[$s(16,7)$]
$\equiv \Theta_{\mathrm{Moon}}=173\aste$,
{\it  \Puna
``Opening of the windows and opening of the court. 
Seeing the 
portal of the ``western side of Thebes''\Egy
where his place is. 
Do not look at the darkness on this day.''}

\item[$s(17,7)$] 
$\equiv \Theta_{\mathrm{Moon}}=185\aste$,
(Figure \ref{figfour}: \SW{Seth})
{\it  \Puna ``Do not speak the name of Seth on this day. 
Who in his lack of knowledge pronounces his name, 
he will not stop fighting in his house of eternity.''}

\end{itemize}

\noindent
According to his calculations $s(16,7)$ was the New Moon day
when one is forbidden to go outside and see the darkness.
The above two unlucky dates are indeed at both sides of point 
Mc $\equiv \Theta_{\mathrm{Moon}}=180\aste$, and
exactly half a lunar cycle away from the luckiest phase at Ma.
This result for the phase angle of the New Moon,
Mc $\equiv \Theta_{\mathrm{Moon}}=180\aste$,
suggests that the epoch $t_0$ of our ephemeris 
of Eq. \ref{Mephe} is correct.
Hence, the Full Moon is at Ma $\equiv \Theta_{\mathrm{Moon}}=0\aste$.

The lucky time points $g(D,M)$ having phase angles $\Theta_{\mathrm{Moon}}$ 
close to $\Theta_{\mathrm{Moon}}=0\aste$ amplify the $P_{\mathrm{Moon}}$ signal.
The lucky points of \SW{Earth} and \SW{Heaven} have the strongest
impact on the $P_{\mathrm{Moon}}$ signal \citep[][\LJnow{p16}]{Jet15}. % RefOK

There are only three lucky time points that mention both 
\SW{Horus} and \SW{Seth}
during the same day (Figure \ref{figthree}: dark squares and open triangles. 
Note that the \SW{Horus} and \SW{Seth}
texts at $\Theta_{\mathrm{Moon}}=73\aste$ in List 2 are from two different days).
The excerpts of these three days are

\begin{itemize}
\item[$g(27,1)$]
$\equiv \Theta_{\mathrm{Algol}}=19\aste$, $\Theta_{\mathrm{Moon}}=278\aste$,
{\it ``Peace between Horus and Seth'' }

\item[$g(27,3)$]
$\equiv \Theta_{\mathrm{Algol}}=38\aste$, $\Theta_{\mathrm{Moon}}=287\aste$,
{\it ``Judgement between Horus and Seth. Stopping the fight'' } 

\item[$g(29,3)$]
$\equiv \Theta_{\mathrm{Algol}}=291\aste$, $\Theta_{\mathrm{Moon}}=312\aste$,
{\it `` The white crown is given to Horus and the red one to Seth'' } 

\end{itemize}

\noindent
In all these three cases, 
Algol is also at its brightest and
the Moon is waxing gibbous, 
since half moon had occurred
at phase angle Md $\equiv \Theta_{\mathrm{Moon}}=270\aste$. 
These texts suggest the reconciliation of the two gods, 
\SW{Horus} and \SW{Seth},
thus peace in Egypt and lucky days, when Algol and 
Moon were simultaneously bright.
The White Crown (Hedjet) represented the kingship of Upper Egypt 
and the Red Crown (Deshret) the rulership of Lower Egypt
\citep[][\LJnow{p321-325}]{Goe01}. % RefOK 

We know from an unrelated text from Edfu that \SW{Horus} 
would benefit from the brightening of the Moon: 
{\it ``When he completes the half month, 
he assumes control of the sky rejuvenated''}
\citep[][\LJnow{p480-482}]{Kap01}. % RefOK
The CC text

\begin{itemize}

\item[$s(26,2)$]  
$\equiv \Theta_{\mathrm{Moon}}=270\aste$,
{\it  \Puna
``Do not lay the foundation of a house. Do not stock a workshop. Do not order any job. Do not do any work on this day. It is day of opening and closing the court and the windows of {Busiris.''}}

\end{itemize}

\noindent
coincides with Md $\equiv \Theta_{\mathrm{Moon}}=270\aste$, 
the first-quarter moon.

At the moment of Full Moon, 
\SW{Horus} was declared ``true of voice'' and ``joyful''; 
related to his victory over \SW{Seth} in the divine 
tribunal \citep[][\LJnow{p480-482}]{Kap01}. % RefOK
The unlucky prognoses of \SW{Seth} also show a connection to $P_{\mathrm{Moon}}$, 
especially to the darker phases of Moon (Figure \ref{figfour}: open triangles).

The only lucky time point of \SW{Horus} overlapping 
the thick curved line in Figure \ref{figone} (Algol's primary eclipse), mentions
\begin{itemize}

\item[$g(28,3)$]  
$\equiv \Theta_{\mathrm{Moon}}=164\aste$,
{\it ``The gods are in jubilation and in joy over 
the making of will for Horus'' } 

\end{itemize}

\noindent
it refers to the will made by \SW{Osiris} that raises
\SW{Horus} to be the ruler of Egypt. Because of this we can not say that the dark phase of Algol
would be always unlucky. 
However, unlucky days follow {\it immediately after} Algol's eclipse
because
the regular distribution of unlucky time points $s(D,M)$
of \SW{Horus}, \SW{Wedjat} and \SW{Sakhmet}
concentrates at $\Theta_{\mathrm{Algol}}=270\aste$ (Figure \ref{figtwo}).

The ``second principle in assigning Lucky and Unlucky Days'' 
probably was

\begin{itemize}

\item[] principle II: 
{\it Use elements from LE1 and LE2 to indirectly describe Algol's changes
(List 1).}

\end{itemize}

List 1 contains the translated full CC passages mentioning 
\SW{Horus}, \SW{Wedjat} and \SW{Sakhmet}.
The short excerpts below give a compact presentation of these passages.
We indicate the cases with a clear connection to LE1 or LE2.
Our notation ``NC'' means that 
there is no clear connection specifically to either myth.

\begin{itemize}

\item[$g(14,2)$]
$\equiv \Theta_{\mathrm{Algol}}=6\aste$ (Figure \ref{figone}: \SW{Horus})
{\it ``Horus receiving the white crown''} (LE2)

\item[$s(5,8)$]
 $\equiv \Theta_{\mathrm{Algol}}=6\aste$ (Figure \ref{figtwo}: \SW{Horus})
{\it  \Puna
``Horus is proceeding
while Deshret sees his image''} (NC)

\item[$g(1,5)$]
$\equiv \Theta_{\mathrm{Algol}}=13\aste$ (Figure \ref{figone}: \SW{Wedjat},
\SW{Sakhmet})
{\it ``Re ... Sakhmet, ... pacify the Wedjat'' } (LE1) 

\item[$g(19,12)$]
$\equiv \Theta_{\mathrm{Algol}}=13\aste$ (Figure \ref{figone}: \SW{Horus})
{\it ``This eye of Horus has come, is complete, is uninjured''} (LE1)

\item[$g(27,1)$]
$\equiv \Theta_{\mathrm{Algol}}=19\aste$ (Figure \ref{figone}: \SW{Horus})
{\it ``Peace between Horus and Seth''} (LE2)

\item[$g(24,3)$]
$\equiv \Theta_{\mathrm{Algol}}=19\aste$ (Figure \ref{figone}: \SW{Horus})
{\it ``giving his throne to his son Horus''} (LE2)

\item[$g(1,7)$]
$\equiv \Theta_{\mathrm{Algol}}=32\aste$ (Figure \ref{figone}: \SW{Horus})
{\it ``entering into heaven. The two banks of Horus rejoice''} (NC)

\item[$g(27,3)$]
$\equiv \Theta_{\mathrm{Algol}}=38\aste$ (Figure \ref{figone}: \SW{Horus})
{\it ``Judgement between Horus and Seth. Stopping the fight''} (LE2)

\item[$g(15,11)$]
$\equiv \Theta_{\mathrm{Algol}}=38\aste$ (Figure \ref{figone}: \SW{Horus}) 
{\it ``Horus hears your words in front of every god'' } (NC)

\item[$g(1,9)$]
$\equiv \Theta_{\mathrm{Algol}}=51\aste$ (Figure \ref{figone}: \SW{Horus}) 
{\it ``Feast of Horus, the son of Isis'' } (NC)

\item[$g(3,2)$]
$\equiv \Theta_{\mathrm{Algol}}=57\aste$ (Figure \ref{figone}: \SW{Wedjat})   
{\it ``Re ... gave the inscription of pacification of Wedjat-eye''} (LE1)

\item[$g(7,9)$]
$\equiv \Theta_{\mathrm{Algol}}=88\aste$ (Figure \ref{figone}: \SW{Horus})
{\it ``followers of Horus ... in the foreign land''} (LE1 or LE2)

\item[$g(28,3)$]
$\equiv \Theta_{\mathrm{Algol}}=164\aste$ (Figure \ref{figone}: \SW{Horus})
{\it ``the making of will for Horus''} (LE1 or LE2)

\item[$g(1,10)$]
$\equiv \Theta_{\mathrm{Algol}}=240\aste$
(Figure \ref{figone}: \SW{Horus}) 
{\it ``Horus... Osiris... Chentechtai... Land''} (NC)

\item[$s(26,1)$]
$\equiv \Theta_{\mathrm{Algol}}=253\aste$ (Figure \ref{figtwo}: \SW{Horus})
{\it \Puna ``day of fighting between Horus and Seth''} (LE2)

\item[$s(11,11)$]
$\equiv \Theta_{\mathrm{Algol}}=253\aste$ (Figure \ref{figtwo}: \SW{Horus})
{\it  \Puna
``the eye of Horus raging in front of Re''} (LE1) 

\item[$s(10,6)$]
$\equiv \Theta_{\mathrm{Algol}}=259\aste$ (Figure \ref{figtwo}: \SW{Wedjat})  
{\it \Puna
``coming forth of Wedjat''} (LE1)

\item[$s(27,8)$]
$\equiv \Theta_{\mathrm{Algol}}=265\aste$ (Figure \ref{figtwo}: \SW{Sakhmet})
{\it  \Puna
``majesty of Sakhmet violates''} (LE1)

\item[ $g(16,4)$]
$\equiv \Theta_{\mathrm{Algol}}=278\aste$ (Figure \ref{figone}:  \SW{Sakhmet})  
{\it ``day of the feast of Sakhmet and Bastet''} (NC)

\item[$s(13,6)$]
$\equiv \Theta_{\mathrm{Algol}}=278\aste$ (Figure \ref{figtwo}: \SW{Sakhmet})
{\it  \Puna
``arrival of Sakhmet ... ``slaughterer-demons''... loose''} (NC)

\item[$s(7,10)$] 
$\equiv \Theta_{\mathrm{Algol}}=278\aste$ (Figure \ref{figtwo}: \SW{Sakhmet})
{\it \Puna {````}slaughterer-demons''\Egy of Sakhmet''} 
(NC)

\item[$g(23,7)$]
$\equiv \Theta_{\mathrm{Algol}}=291\aste$ (Figure \ref{figone}: \SW{Horus}) 
{\it ``Feast of Horus in Athribis''} (NC)

\item[$g(29,3)$]
$\equiv \Theta_{\mathrm{Algol}}=291\aste$ (Figure \ref{figone}: \SW{Horus}) 
{\it ``white crown is given to Horus ... red one to Seth''} (LE2)

\item[$s(20,9)$]
$\equiv \Theta_{\mathrm{Algol}}=291\aste$ (Figure \ref{figtwo}: \SW{Horus})
{\it \Puna   ``angered on the island ...
inspected by ... Horus.''} (NC)

\item[$g(9,5)$] 
$\equiv \Theta_{\mathrm{Algol}}=303\aste$
(Figure \ref{figone}: \SW{Sakhmet})
{\it ``gods are joyful over the matter of Sakhmet''} (LE1)

\item[$g(30,10)$]
$\equiv \Theta_{\mathrm{Algol}}=303\aste$ (Figure \ref{figone}: \SW{Wedjat})  
{\it ``coming forth of Shu to bring back Wedjat''} (LE1)

\item[$g(29,5)$] 
$\equiv \Theta_{\mathrm{Algol}}=309\aste$
(Figure \ref{figone}: \SW{Sakhmet})
{\it ``sole mistress Sakhmet the great ... gods are pleased''} (NC)

\item[$g(18,1)$]
$\equiv \Theta_{\mathrm{Algol}}=322\aste$ (Figure \ref{figone}: \SW{Horus})
{\it ``magnifying the majesty of Horus over his brother''} (LE2)

\item[$g(6,9)$]
$\equiv \Theta_{\mathrm{Algol}}=322\aste$ (Figure \ref{figone}: \SW{Wedjat}) 
{\it ``they catch Wedjat together with their followers''} (LE1)  

\end{itemize}

The lucky prognoses of the days related to the main protagonists of LE1,
\SW{Horus}, \SW{Wedjat} and \SW{Sakhmet}, 
had a strong impact on the $P_{\mathrm{Algol}}$ 
signal \citep[][\LJnow{p8}]{Jet15}. % RefOK
All time points of these three deities
have extremely similar $\Theta_{\mathrm{Algol}}$ distributions
(Figures \ref{figone} and \ref{figtwo}).
The lucky prognoses are centered at Aa  at $\Theta_{\mathrm{Algol}}=0\aste$
in the middle of the brightest phase of Algol (Figure \ref{figone}). 
The unlucky prognoses of these three SWs 
concentrate close to Ad at $\Theta_{\mathrm{Algol}}=270\aste$,
immediately after Algol's eclipse (Figure \ref{figtwo}). 

The passages between $\Theta_{\mathrm{Algol}}=6\aste$ and $51\aste$ describe 
feasts or peaceful actions by \SW{Horus}, \SW{Wedjat} and \SW{Sakhmet}. 
Those regarding the crowning or judgement of \SW{Horus} are related to LE2.
The bringing back of \SW{Wedjat} is related to LE1,
as well as the pacifications of \SW{Sakhmet} or \SW{Wedjat}.
The only exception is {\it  \Puna
``Horus is proceeding while Deshret sees his image''}
at $\Theta_{\mathrm{Algol}}=6\aste$ which might not be directly related 
to either LE1 or LE2.
Notably, this unlucky date close to point Aa 
is the only unlucky time point that deviates from the
other seven unlucky time points concentrated
close to Ad in Figure \ref{figtwo}.

The text at $\Theta_{\mathrm{Algol}}=57\aste$ refers to the order 
that \SW{Re} gave in LE1 to save mankind from the wrath 
of \SW{Wedjat} (Eye of \SW{Horus}), synonymous with \SW{Sakhmet}.
As Algol's eclipse is approaching at $\Theta_{\mathrm{Algol}}=88\aste$,
\SW{Horus} enters {\it ``the foreign land''} (LE1 or LE2).
His {\it ``will is written''} at $\Theta_{\mathrm{Algol}}=164\aste$ (LE1 or LE2).
This particular time point $g(28,3)$
coincides with the moment when Algol's eclipse 
can be observed with naked eye (Figure \ref{figone}: thick curved line at Ac). 
 
The interval between $\Theta_{\mathrm{Algol}}=253\aste$ 
and $\Theta_{\mathrm{Algol}}=291\aste$ 
is filled with descriptions of anger and aggression by 
\SW{Horus}, \SW{Wedjat} and \SW{Sakhmet},
or contains hostile elements such as the {\it ``slaughterer-demons''}.
These specific demons were believed to punish mankind 
on behalf of \SW{Sakhmet} and 
were considered the cause of diseases and symptoms whose pathology 
was not well understood \citep{Luc10}. % RefOK
Some texts are clearly related to the
transformation of \SW{Wedjat} into the raging \SW{Sakhmet} (LE1).
Two out of the three prognoses at $\Theta_{\mathrm{Algol}}=291\aste$, are lucky. 
After these three, all remaining texts are lucky, 
\textit{i.e.} there are only good prognoses in the end of Algol's cycle. 
This suggests the pacification of \SW{Sakhmet}, as in LE1.
{\it `` The coming forth of Shu to bring back Wedjat''}
is mentioned at $\Theta_{\mathrm{Algol}}=303\aste$ (LE1).
\SW{Shu} succeeds in his intention, 
{\it `` they catch Wedjat together with their followers''},
at $\Theta_{\mathrm{Algol}}=322\aste$ (LE1).
Then, the same stories begin anew at $\Theta_{\mathrm{Algol}}=6\aste$.

In short, all the texts of \SW{Wedjat} and \SW{Sakhmet}
that were not discussed earlier \citep{Jet15}, % RefOK
also support the idea that principle II 
was used to assign the prognoses of CC.
The order of List 1 texts connected to LE1 
more or less follows the plot of LE1,
but the order of texts connected to LE2 does not follow the plot of LE2.
This is not unexpected, 
because several events are related to
\SW{Seth}, who is not connected to the $P_{\mathrm{Algol}}$ 
signal \citep{Jet15}. % RefOK

The ``third principle in assigning 
Lucky and Unlucky Days'' 
could have been

\begin{itemize}

\item[] principle III:
{\it Use LE2 for indirect description of the lunar phases (List 2).}

\end{itemize}

The unlucky prognosis texts of \SW{Seth} and \SW{Osiris} support this idea.
These prognoses are very clearly concentrated to the dark phases of the Moon 
(Figure \ref{figfour}: open triangles and closed circles).
The descriptions from List 2 are

\begin{itemize}

\item[$s(12,2)$] 
$\equiv \Theta_{\mathrm{Moon}}=100\aste$
(Figure \ref{figfour}: \SW{Seth})
{\it  \Puna
``His head, who did rebel against his lord, is cut off''}

\item[$s(13,3)$] 
$\equiv \Theta_{\mathrm{Moon}}=117\aste$
(Figure \ref{figfour}: \SW{Seth} and \SW{Osiris})
{\it  \Puna
``day of severing''}

\item[$s(14,3)$] 
$\equiv \Theta_{\mathrm{Moon}}=129\aste$
(Figure \ref{figfour}: \SW{Osiris})
{\it  \Puna 
``gods are sad over the action against Osiris' 
place''}

\item[$s(11,12)$]
$\equiv \Theta_{\mathrm{Moon}}=137\aste$
(Figure \ref{figfour}: \SW{Seth}) 
{\it  \Puna
``repelled the confederacy of Seth to the eastern desert''}

\item[$s(14,5)$] 
$\equiv \Theta_{\mathrm{Moon}}=139\aste$
(Figure \ref{figfour}: \SW{Osiris})
{\it  \Puna
``Weeping of Isis and Nephthys''}

\item[$s(18,3)$] 
$\equiv \Theta_{\mathrm{Moon}}=178\aste$
(Figure \ref{figfour}: \SW{Seth})
{\it  \Puna
``tumult by the children of Geb: Seth and his sister Nephthys'''}

\item[$s(17,7)$] 
$\equiv \Theta_{\mathrm{Moon}}=185\aste$
(Figure \ref{figfour}: \SW{Seth})
{\it  \Puna
``Do not speak the name of Seth on this day''}

\item[$s(17,8)$] 
$\equiv \Theta_{\mathrm{Moon}}=190\aste$
(Figure \ref{figfour}: \SW{Seth})
{\it  \Puna
``The going of Seth, ... they repelled his followers''}

\item[$s(19,4)$] 
$\equiv \Theta_{\mathrm{Moon}}=195\aste$
(Figure \ref{figfour}: \SW{Osiris})
{\it  \Puna
``Making of ointment for Osiris''}

\item[$s(20,2)$] 
$\equiv \Theta_{\mathrm{Moon}}=197\aste$
(Figure \ref{figfour}: \SW{Seth})
{\it  \Puna
``The rebels against their lord were overthrown''}

\item[$s(26,1)$] 
$\equiv \Theta_{\mathrm{Moon}}=266\aste$
(Figure \ref{figfour}: \SW{Horus} and \SW{Seth})
{\it  \Puna
``day of fighting between Horus and Seth''}

\item[$s(24,8)$] 
$\equiv \Theta_{\mathrm{Moon}}=275\aste$
(Figure \ref{figfour}: \SW{Seth})
{\it  \Puna
``Do not pronounce the name of Seth''}

\end{itemize}

Seth's rebellion is first referred to at $\Theta_{\mathrm{Moon}}=100\aste$.
It is followed by a long story of { \Puna ``the day of severing''}
on $s(13,3)$.
On the next day $s(14,3)$,  { \Puna ``gods are sad over the
action against Osiris's place''}.
\SW{Seth} is repelled in a tumult ($\Theta_{\mathrm{Moon}}=137\aste$).
The goddesses Isis and Nephthys are mourning the death of \SW{Osiris}  
($\Theta_{\mathrm{Moon}}=139\aste$).
\SW{Seth} causes another tumult ($\Theta_{\mathrm{Moon}}=178\aste$).
CC advises not to pronounce 
his name at New Moon ($\Theta_{\mathrm{Moon}}=185\aste$).
His influence begins to wane ($\Theta_{\mathrm{Moon}}=190\aste$).
{\it  \Puna ``Making of ointment for Osiris''} follows 
at $\Theta_{\mathrm{Moon}}=195\aste$.
\SW{Seth} is overthrown and judged ($\Theta_{\mathrm{Moon}}=197\aste$).
Yet, \SW{Seth} and \SW{Horus} continue 
their fight ($\Theta_{\mathrm{Moon}}=266\aste$). 
In the end, CC advises not to pronounce the name of \SW{Seth}
($\Theta_{\mathrm{Moon}}=275\aste$).
All these texts are connected {\it only} to LE2.
They mostly follow the plot of LE2 in the order of $\Theta_{\mathrm{Moon}}$,
but there are contradictions. 
For example, the beginning of the cycle would
be the expected place for
the fight at $\Theta_{\mathrm{Moon}}=266\aste$.

\SW{Osiris} is also connected to LE2 in the CC.
Therefore, it is logical that the unlucky texts of \SW{Osiris} 
were connected to $P_{\mathrm{Moon}}$ \citep[][\LJnow{p18}]{Jet15}. % RefOK

The three previously discussed lucky prognoses 
on $g(27,1), g(27,3)$ and $g(29,3)$ mentioning
both \SW{Seth} and \SW{Horus} show a reconciliation (LE2) when 
Algol was at its brightest and
the Moon was waxing gibbous
after Md $(\Theta_{\mathrm{Moon}}=270\aste)$.

It would not be logical to study the texts 
of \SW{Seth} in connection with LE1,
since he is not one of the protagonists in the myth.
\SW{Horus} personally plays a part only in LE2,
but his Eye is a protagonist in LE1.
Neither the lucky, nor unlucky, prognoses of \SW{Horus} were connected 
to $P_{\mathrm{Moon}}$ \citep[][\LJnow{p7}]{Jet15}, % RefOK
\textit{i.e.} they were randomly distributed as 
a function of $\Theta_{\mathrm{Moon}}$
(Figures \ref{figthree} and \ref{figfour}: closed squares).

Because \SW{Horus} is present as a name in both LE1 and LE2, the 
same texts mentioning \SW{Horus}
are repeated when arranged in the order of increasing $\Theta_{\mathrm{Algol}}$
(List 1)
and when arranged in the order of increasing $\Theta_{\mathrm{Moon}}$
(List 2).
This means that we have 
(or the scribes had) a choice to interpret those texts either in 
relation to the cycle of Algol or the cycle of the Moon.
We do not know if the scribes faithfully 
repeated always the same mythological texts in relation to
the cycle of Algol, the cycle of the Moon or both of these cycles.
However, if they did,
these \SW{Horus} texts may have a double relation.

The 1st example of a double relation to both Algol and Moon, 
or to LE1 and LE2, is the text

\begin{itemize}

\item[$g(28,3)$] 
{\it ``The gods are in jubilation and in joy over 
the making of will for Horus, son of Osiris, 
to pacify Onnophris in the underworld.''}

\end{itemize}

\noindent
At $\Theta_{\mathrm{Algol}}=164\aste$,
it can describe the return of the lost Eye of Horus
during the eclipse of Algol (LE1),
but more likely it refers to the will made by \SW{Osiris} that raises
\SW{Horus} to be the ruler of Egypt at 
$\Theta_{\mathrm{Moon}}=300\aste$ (LE2). 
It is also related to the time of gestation and infancy of 
\SW{{Horus}}, 
when he was hidden from enemies' sight by his mother Isis. All of 
these three alternative interpretations
symbolize the cyclic rejuvenation of royal power over Egypt.

The 2nd example is the text
\begin{itemize}

\item[$g(7,9)$]
{\it ``The crew and followers of Horus have assembled in the foreign land,
to make known that Horus smites him who rebels against his lord.'' }

\end{itemize}

\noindent
On this day, the phase angles of Algol and the Moon are nearly equal 
$\Theta_{\mathrm{Algol}}=88\aste$ and $\Theta_{\mathrm{Moon}}=73\aste$.
The smiting of the rebels is equally applicable to punishing 
the mankind for its wicked ways (LE1)
and to the battle against \SW{Seth} with his followers (LE2).

However, only the unlucky texts of \SW{Seth} and \SW{Osiris}
clearly follow the principle III,
and perhaps the three lucky texts mentioning both \SW{Horus} and \SW{Seth}.
Many of the remaining \SW{Horus}, \SW{Seth} or \SW{Osiris} texts may
have more complicated connections to LE1 and LE2,
or to other myths.

We can confirm principle I which 
connects the lucky prognoses to the moments in the middle of the bright 
phases of Algol and the Moon. 
More tentatively, we present principles II and III,
\textit{i.e.} a specific connection of the LE1 and LE2 myths 
to the phases of Algol and the Moon.
The hemerological tradition had already existed for centuries,
if not millennia, and accordingly had accumulated into itself cultural 
layers from different historical eras. 
However, this uncertainty in {\it principles} does not alter our main result 
that the scribes connected the {\it texts} of the two legends LE1 and LE2
to the phases of Algol and the Moon.

Argument VIII: The texts of List 1 ($\Theta_{\mathrm{Algol}}$ order) 
and List 2 ($\Theta_{\mathrm{Moon}}$ order)
show that the LE1 and LE2 legends
could have been used to describe indirectly the 
regular changes of Algol and the Moon.

\subsection{Rejuvenation and kingship}

Rejuvenation, the power to disappear and reappear,
was associated with \SW{Horus}.

Everything in the ancient Egyptian worldview was 
repeating in a cyclic manner: the sky, the celestial phenomena, 
the Nile, the winds, the clouds,
the migrating birds and fish, 
all life in its individual or 
holistic sense \citep[][\LJnow{p452-459}]{Lei94}. % RefOK
Even the highest power in Egypt, 
the divine king, was subject to a continuous cycle,
first representing \SW{Horus} on Earth and then \SW{Osiris} in afterlife.
The ancient Egyptian deities were subject to the same eternal 
rules of recurrence.

The discovery of Algol's variability might have first 
astonished the scribes.
Had the cause of variability been non-repeating, such as supernova, 
they would have been baffled by it, but it must have been reaffirming to
them to notice the regularity of Algol's variability, 
well suited to their worldview: the idea of the continuous 
struggle between the forces of chaos and order 
\citep[][\LJnow{p2}]{Mag13}. % RefOK
It must have taken some effort 
to incorporate this new phenomenon into 
their religion and mythology, 
but evidently they were able to do that.

It was probably of utmost importance to the scribes
to determine 
the period of this phenomenon,
because this would have allowed them 
to interpret correctly the divine events 
relating to it and also to incorporate 
these events into their explanation of cosmos.

We must remember that the ancient Egyptians did not 
practice natural sciences in the modern sense, 
but expressed their worldview and 
all observable phenomena 
in the context of religion and myths 
\citep[][\LJnow{p188}]{Lie00}. % RefOK
 Many of their discoveries 
were known only by the experts of
 religion and magic, the scribes, 
who could interpret the indirect mythological descriptions 
of the observed phenomena.
The intention of the writers of CC was not necessarily to prevent 
outsiders
from understanding the connections to Algol and the Moon, 
but for three millennia their indirect mythological references 
have hidden from sight the basic periodic 
principles of assigning
Lucky and Unlucky Days (\textit{i.e.} principles I, II and III). 

The idea of rejuvenation was important in ancient Egyptian mythology.
The Moon was used as a symbol of rejuvenation, 
called ``the one that repeats its form'' 
\citep[][\LJnow{p480-482}]{Kap01}. % RefOK 
Likewise, a vanishing and reappearing star 
would have been suggestive of the 
restoration of the eye of \SW{Horus} 
alongside his kingship 
\citep[][\LJnow{p6}]{All05},
\citep[][\LJnow{p233-240}]{Edw95} and
\citep[][\LJnow{p132-133}]{Tro89}, % RefOK
and reaffirming 
of the Egyptians' cyclic worldview. 
The eyes of \SW{Horus} were associated with the crowns of the kings, 
as 
symbols of sovereignty \citep[][\LJnow{p476-480}]{Gri01}. % RefOK
Hence, a regularly variable star would naturally 
be linked with the divine cycles (\SW{Horus} and his eye), 
much like the Moon was seen to have 
rejuvenative power \citep[][\LJnow{p480-482}]{Kap01}. % RefOK
Algol's eclipses could have been 
considered the blinding of the eye of \SW{Horus} by \SW{Seth}, 
but on the other hand the return of the Eye of \SW{Horus} 
would have reaffirmed the restorative, 
life-giving powers of the gods and kings of Egypt.
\SW{Horus} or his eye are supposed 
to have been linked with varying celestial 
objects depending on the context. 
\citet[][\LJnow{p137-139}]{Kra16} interpreted 
the Eye of Horus in CC to be Venus because the narration regarding 
it can also be taken to mean the yearly absence of Venus from the 
night sky as it is transformed from a morning star into an evening star. 
The dramatic changes in the brilliancy of Venus would have been 
indicative of divine fighting, injuring and rivalry. Additionally, 
Krauss suggests that the stars called {\it sehed} in the Pyramid Texts 
such as Horus are planets because the text attributes to them 
the ability to move freely but could this also include the ability 
to vanish from the sky? 
It is logical that Algol would have been represented 
in mythological texts as \SW{Horus} 
or his eye, as were most of the 
planets in the known 
Egyptian astronomical texts \citep[][\LJnow{p114}]{Cla95}. % RefOK 
Considering the significance of the name \SW{Horus} to the periodic signal, 
the ancient Egyptians could have
believed in some connection between Algol, royal power and the cosmic order.

Argument IX:
Algol could have been 
naturally associated with \SW{Horus} and called as such, 
because Algol can disappear and reappear.

\subsection{Astrophysical evidence \label{astrophys}}

The period of Algol must have been shorter three millennia 
ago \citep{Kwe58,Bie73,Sod75}. % RefOK

It is a fascinating idea that the ancient Egyptians 
would have
discovered Algol's variability 
over three thousand years ago, 
noticed its regularity,
determined its period and incorporated
this phenomenon into their mythology.
Our modern interpretations of their concepts are mostly circumstantial.
The interpretation of ancient Egyptian texts is complex
when the phenomena are not
everyday concrete matters such as agriculture, climate,
weather, time keeping, geography or geometry.
Although the CC deals with astronomy only indirectly,
it contains evidence that the scribes made recordings of
a concrete phenomenon later discovered by modern natural science:
the regular changes of the eclipsing binary Algol. 
Several astronomical and astrophysical 
considerations \citep{Jet13,Jet15} % RefOK
support the idea that their prolonged naked eye observations 
revealed the same discoveries of Algol
that Montanari (variability) and Goodricke (regular variability) 
made about three millennia later.

Naked eye can discover regular variability in the Sun, the Moon,
the planets and the variable stars. 
\citet{Jet13} % RefOK
formulated eight astronomical criteria which
showed that only the periods of the Moon and Algol could be discovered
from CC, and it was exactly these two periods that they rediscovered.
Their period analysis also showed that 2.850 days is 
the strongest real periodicity governing the assignment 
of the lucky prognoses in CC, after the lucky prognoses 
connected to the synodic lunar month are removed. 
The mass transfer from Algol~B to Algol~A 
is a well established phenomenon \citep{Sod75,Sar93}. % RefOK
This mass transfer should cause a period increase
\citep{Kwe58,Bie73}. % RefOK
Yet, no-one had confirmed the presence of this phenomenon in over 230 years,
since Goodricke determined the period of Algol in 1783.
The period of 2.850 days in CC is 0.017 days shorter 
than the current orbital period of Algol, 2.867 days. 
The mass transfer between Algol B and A 
could have caused this period increase 
during the past three millennia.
The required mass transfer rate 
\citep[][\LJnow{p8}]{Jet13} % RefOK
was in excellent agreement
with the predictions of the best evolutionary model of 
Algol \citep[][\LJnow{p540}]{Sar93}.
Furthermore, it was shown that Algol's inclination has 
remained stable \citep{Zav10,Bar12}, 
\textit{i.e.} eclipses occurred also in that historical era
and the ancient Egyptians 
would have been able to record these events \citep{Jet13}.

Argument X: Astrophysical considerations support
the idea that the 2.850 days period in CC can be the period of Algol.

\subsection{Cultural evidence and lack of it}

Our arguments prove that the ancient Egyptians
{\it could have} recorded Algol's period into CC, 
but the very same arguments do not definitively prove 
that they {\it did} so. 
However, such cultural aspects
are not important for all of our arguments. 
We will first discuss the arguments that pose no problems,
and then those that do.

There is cultural evidence about the ``hour-watchers'', as well as
about the connection between their observations and the religious rituals
(Arguments I and II).
Algol's variability could have been easily observed in any ancient culture,
unless the geographical location of this culture 
prevented observations (Argument III).
Many excerpts in List 1 and List 2 are definitively connected to the
LE1 and LE2 legends (Argument VII). 
The order of these excerpts in our Lists 1 and 2 
is based on statistical analysis, not on cultural aspects.
It is therefore
not accidental that this order makes sense (Argument VIII).
Statistical, astronomical and astrophysical 
evidence \citep{Por08,Jet13,Jet15}
supports our last argument (Argument X).

{\it If} the ``hour-watchers'' noticed Algol's variability
and {\it just} recorded the observed eclipses into CC,
then modern period analysis would detect the 2.850 days 
periodicity (Arguments IV and V).
{\it If} the references to Algol and the Moon in CC are indeed indirect,
it is, and it will be, very difficult to find any 
definitive cultural proof about Argument VI.
Statistical analysis by \citet{Jet15} has revealed a connection 
between the 2.850 days period
and \SW{Horus}, but our cultural
interpretation can be questioned (Argument IX).

We do admit that a specific identification of Algol is missing
(FAQ \ref{nothourstar}) but this is a general problem regarding 
all but a few stars and planets. The Pyramid Texts make it obvious 
that Horus is a star, but the identification of the star is 
a matter of debate. 
Late Period texts identify Horus-son-of-Isis 
as god of the morning (mentioned as a star also in Coffin Texts 
from a much older period, the First Intermediate Period), 
from which  \citet[][\LJnow{p137-141}]{Kra16} 
concluded that Horus-son-of-Isis and 
Haroeris (the elder Horus) are Venus as morning star 
and evening 
{star.
This} does not exclude other interpretations, 
particularly since we know that many other celestial objects 
received the title or association to Horus. 
Descriptions such as 'Horus who ascends as gold 
from upon the lips of the akhet' (Coffin Texts 255), 
are applicable to all but circumpolar stars. 
Krauss analyzed other passages as well 
from Coffin Texts and the Book of the Dead 
which prove the association of the Eye of Horus 
to something else than the Sun or the Moon. 
It would be worthwhile to also study 
the celestial diagrams of the Late Period 
which include Greek and Mesopotamian influences, 
to discover in greater detail the connection 
of the Greek constellation of Perseus to the ancient Egyptian star names.

\section{Conclusions}
\setlist[itemize]{topsep=5pt,leftmargin=25pt}
\setlist[enumerate]{topsep=5pt, leftmargin= 25pt}

We have presented ten arguments 
which show that the ancient Egyptian scribes,
the ``hour-watchers'',
had the possible
means and the motives for recording the period of Algol in CC. 
Those arguments are combined here.

\begin{itemize}

\item[] Argument I:
For thousands of years, the ``hour-watchers'' practiced the tradition
of timekeeping by observing hour-stars.
If Algol was not an hour-star, it certainly 
belonged to some hour-star pattern or related constellation.

\end{itemize}

\noindent
The scribes
observed regularly about 70 bright stars,
or most probably a lot more,
in a region where there are about 300 clear nights every year.
They practiced this tradition for the timing of religious rituals.

\begin{itemize}

\item[] Argument II:
Proper timing of the nightly religious rituals
relied on the fixed hour-star patterns.

\end{itemize}

\noindent
The ``hour-watchers'' probably discovered Algol's variability 
from the changes that its eclipses caused 
in its hour-star pattern.

\begin{itemize}

\item[] Argument III: A naked eye can easily
discover the significant hour-star pattern
change caused by Algol's eclipse.

\end{itemize}

\noindent
These changes followed the 
regular ``3+3+16=19'' and ``19+19+19=57'' days cycles. 

\begin{itemize}

\item[] Argument IV: The scribes could have discovered Algol's 
$2.850=57/20$ days period
from long-term observations of 
the regular 19 and 57 days eclipse cycles.

\end{itemize}

\noindent
Three alternative methods could 
have been used in recording the 2.850 days period.

\begin{itemize}

\item[] Argument V:
 The ancient Egyptian scribes may have calculated
the $57/20=2.850$ days period of Algol 
from long-term observations (1st method).
They may not have calculated this 2.850 days period,
because the 19 days and 57 days cycles already 
perfectly predicted {\it all} night-time eclipses of Algol (2nd method), 
or they may have just recorded the observed night-time eclipses into CC
(3rd method).

\end{itemize}

\noindent
The scribes did not describe Algol's regular changes directly.

\begin{itemize}

\item[]
Argument VI: To avoid violating cosmic order,
the scribes would have referred to Algol's changes only indirectly.

\end{itemize}

\noindent
Two legends, 
``the Destruction of Mankind'' and ``the Contendings of Horus and Seth'', 
could have been used
to indirectly describe the changes of Algol and Moon.

\begin{itemize}

\item[]
Argument VII: Even a quick glance on
List 1 ($\Theta_{\mathrm{Algol}}$ order)
and
List 2 ($\Theta_{\mathrm{Moon}}$ order)
reveals that numerous CC texts are excerpts
from the LE1 and LE2 legends.

\end{itemize}

\noindent
These two legends were probably used to
describe celestial phenomena as activity of gods.

\begin{itemize}

\item[]
Argument VIII: The texts of List 1 ($\Theta_{\mathrm{Algol}}$ order) 
and List 2 ($\Theta_{\mathrm{Moon}}$ order)
show that the LE1 and LE2 legends
could have been used to describe indirectly the 
regular changes of Algol and the Moon.

\end{itemize}

\noindent
Algol would have been considered as a manifestation of \SW{Horus}.

\begin{itemize}

\item[]
Argument IX:
Algol could have been 
naturally associated with \SW{Horus} and called as such, 
because Algol can disappear and reappear.

\end{itemize}

\noindent
We have presented evidence that the period of Algol in CC
was 0.017 days shorter than today \citep{Por08,Jet13,Jet15}.

\begin{itemize}

\item[]
Argument X: Astrophysical considerations support
the idea that the 2.850 days period in CC can be the period of Algol.

\end{itemize}

\noindent
It is not even necessary to present this many arguments 
to convince the reader that it would be more complicated 
to explain the statistically significant and accurate
$2.850 \pm 0.002$ days period with something else than Algol. 
No-one disputes that Algol's variability is easy to observe. 
The lack of other reasonable alternatives is also a part of the evidence. 
It is not only that Algol is the simplest explanation 
but so far no-one has been able to think of any other 
reasonable alternative explanation for the 2.850 days periodicity in CC.
If no-one can answer our last Frequently Asked Question 
(Appendix B: FAQ \ref{ourquestion}),
then the following famous aphorism by the Italian Nobel Prize winner
Luigi Pirandello must be true:
a thing {\it ``is so if it seems so''} \citep{Ben86}.

\begin{acknowledgement}
We dedicate this paper to the memory of our two co-authors and friends,
PhD Jaana Toivari-Viitala (May 16th, 1964 - May 12th, 2017)
and
prof. Tapio Markkanen (Jan 27th, 1942 - Aug 28th, 2017).
We acknowledge
PhD Patricia Berg,                
Dr. Robert J. Demaree, 
PhD Heidi Jauhiainen,
prof. Karri Muinonen and prof. Heikki Oja 
for their comments
of the text of this article. 

\end{acknowledgement}

%\bibliography{porceddu} 

\clearpage

\appendix

\section{List of translations to Ancient Egyptian language}

\begin{center}
\includegraphics[scale=0.9]{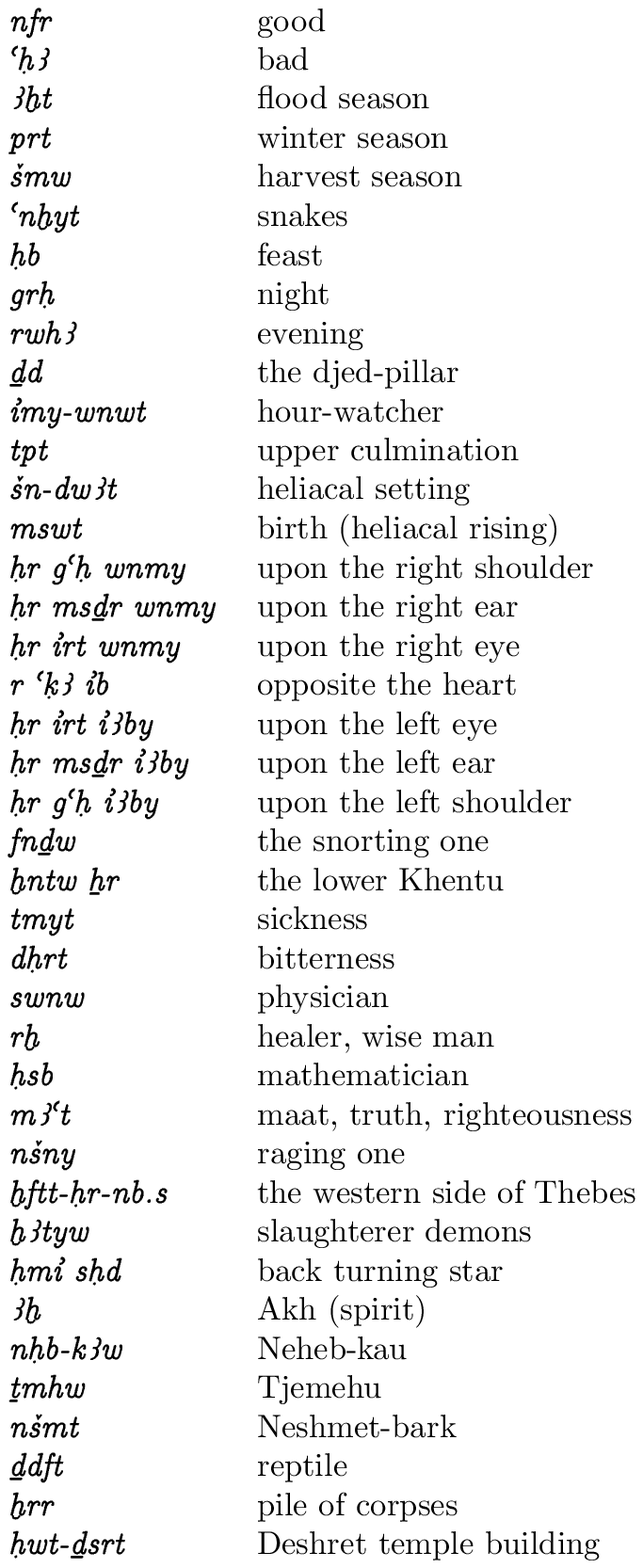}
\end{center}

\newpage

\section{Frequently asked questions (FAQ)}

% ============================================================
% Revise begins =====================================
% ============================================================
We have rewritten innumerable versions of this manuscript.
The list below contains some of the most frequently asked questions (FAQ)
about our research, as well as our short answers.
Some of these questions have arisen from misunderstandings of our research.
However, we want to answer also those questions, because it was impossible 
to foresee them coming, or to prepare answers to them beforehand.
The sections and the paragraphs (par),
where our more detailed answers can be found,
are given in parenthesis.
These answers are aimed to ensure that our research 
better stands the test of time and sceptics.
The last question \ref{ourquestion}
is our own.
So far, we have not received a reasonable alternative 
answer to this simple question.

\begin{enumerate} 

\item \label{deitydate}
Several CC passages mention different deities,
but not all of them are included in your analysis. 
How do you know which one of them determines the calendar date?
\answer{The deities do not determine the dates.
The dates are known (Eqs. \ref{edays} and \ref{timepoints},
Sect. \ref{materials}: 4th and 6th par,
 Sect. \ref{methods}: 4th last par).}

\item \label{dating}
Your statistical analysis is flawed, because the exact
historical dating of CC is uncertain.
\answer{The results of our statistical analysis
do not depend on the zero epoch in time. 
The time points within CC are unambiguously defined and computed
(Eqs. \ref{edays} and \ref{timepoints}).
Hence, the results of our statistical analysis
would be the same if CC was dated to
ice age, stone age or a million years ahead into the future
(Sect. \ref{methods}: 4th last par).}

\item \label{smallsamples}
Your samples are too small to allow reliable statistical conclusions.
\answer{The ephemerides of Eqs. \ref{Aephe} and \ref{Mephe} are based
on large samples of over 500 time points. We use these ephemerides to
rearrange the CC passages into the unambiguous order of Lists 1 and 2.
These ephemerides work like an accurate 
modern clock when studying Lists 1 and 2.
Furthermore, the binomial significance estimates \citep{Jet15} % RefO
$Q_{\mathrm{B}}$ were reliable even for small samples
(Sect. \ref{methods}: 3rd last par).}

\item \label{amiss}
Something must be seriously amiss in your method
if a term like "Earth" has a strong correlation 
with Algol's period.
\answer{The ephemeris of Eq. \ref{Aephe} reveals what
the authors of CC wrote when they were observing Algol 
at its different phases.
Nothing is ``amiss'' in our method,
unless those CC texts were not written on Earth  
(Sect. \ref{methods}: 3rd last par). 
This is natural because feast dates 
where often described as feasts in \SW{Earth} and in \SW{Heaven}.}

\item \label{translation}
Your CC translations are incomplete.
\answer{We have used the hieroglyphic transcription
of \citet{Lei94} % RefOK
and his original photos of CC,
as well the translations of \citet{Bak66}  
(in English) and \citet{Lei94} % RefOK
(in German).
If some translation modifications could be shown to be necessary,
this would not change the general course of events in Lists 1 and 2,
or the SW identifications (Sect. \ref{methods}: 2nd last par).}

\item \label{chitest}
You should establish the statistical validity of your 
analysis by applying a $\chi^2$-test to your samples.
\answer{Unfortunately, this is not possible for our data samples
(Sect. \ref{methods}: last par). }

\item \label{nothourstar}
A definitive identification of 
Algol 
in the Decanal or "star clocks" lists is missing. 
This weakens your hypotheses.
\answer{It is difficult to identify 
Egyptian star names in general, but
such a definitive identification is not needed, 
because Algol caused easily observable 
{changes in 
hour-star} patterns 
or related constellations whether or not it was an actual 
hour-star
(Sect. \ref{hourstars}: 4th-7th par, Sect. \ref{rituals}: 5th par).}

\item \label{speculative}
Most of your presented arguments are speculative, 
and thus your claims are vague and unproven.
Your presented hypotheses are far from being proved.
\answer{This shifts the argumentation from the specific to the general. 
We make no hypotheses.
The previous studies have confirmed the presence 
of the extremely significant 2.850 days period in 
CC \citep{Jet13,Jet15}.
Here, we formulate ten specific Arguments I-X in Sects. 
\ref{hourstars}-
\ref{astrophys}.
It is easy make this type of
general and subjective statements about our research
without presenting any evidence against even one of our ten arguments.}

\item \label{primaryresearch}
Your manuscript does not report primary research.
\answer{We show that the scribes had the possible
means and motives to
write the descriptive texts of the regular changes
of Algol and the Moon into CC.
This must be primary research, 
unless someone else has solved this question before us.}

\item \label{wrongjournal}
The scope of your research is unsuitable for this journal.
\answer{This journal 
``disseminates research in both observational and 
theoretical astronomy, ... 
as well the surveys dedicated to astronomical history and education.''}

\item \label{ourquestion}
If the significant 2.850 days period in CC is not connected to Algol,
then the following question made by \citet{Jet15} 
needs to be addressed:
``What was the origin of the phenomenon that occurred every third day, 
but always 3 hours and 36 minutes earlier than before, 
and caught the attention of Ancient Egyptians?''
In other words,
what happened three times in a row
at the night-time? Then it occurred during
the daytime? After a gap of 13 days, it occurred again during the night-time?

\end{enumerate}

\section{List 1}
\setlist[itemize]{topsep=5pt,leftmargin=45pt}
\setlist[enumerate]{topsep=5pt, leftmargin= 45pt}
\begin{itemize}
\item[$g(14,2)$]
$\equiv \Theta_{\mathrm{Algol}}=6\aste$ (Figure \ref{figone}: \SW{Horus})
{\it
``The day of the majesty of Horus receiving the white crown. 
His Ennead is in a great celebration. 
Make offering to the gods of your city. 
Pacify your ``akh''\Egy.''}

\item[$s(5,8)$]
 $\equiv \Theta_{\mathrm{Algol}}=6\aste$ (Figure \ref{figtwo}: \SW{Horus})
{\it  \Puna
``The majesty of Horus is proceeding
while Deshret sees his image. 
Every approach to him is met with rage.''}

\item[$g(1,5)$]
$\equiv \Theta_{\mathrm{Algol}}=13\aste$ (Figure \ref{figone}: \SW{Wedjat} 
and \SW{Sakhmet})

{\it ``Doubled are the offerings, presented the ritual foods. 
It is the ``Neheb-kau''\Egy feast 
of the gods in front of Ptah, 
in Ta-tenen in all the temples of the gods and goddesses, in front of Re and his followers. He himself is surrounded by Ptah-Sokar, Sakhmet, Nefertem,
Horus-Hekenu and Maahes, the son of Bastet. Light a great fire, pacify the Wedjat. It will be good on this day.'' 
} 

\item[$g(19,12)$]
$\equiv \Theta_{\mathrm{Algol}}=13\aste$ (Figure \ref{figone}: \SW{Horus})
{\it 
``Feast for your god! Propitiate your ``akh'', 
because this eye of Horus has come, is complete, 
is uninjured and there is no claim against it.''}

\item[$g(27,1)$]
$\equiv \Theta_{\mathrm{Algol}}=19\aste$ (Figure \ref{figone}: \SW{Horus})
{\it  
``Peace between Horus and Seth. Do not kill any snakes on this day. 
Make a good day!''}

\item[$g(24,3)$]
$\equiv \Theta_{\mathrm{Algol}}=19\aste$ (Figure \ref{figone}: \SW{Horus})
{\it
``The arrival of Isis joyful and Nephthys 
rejoicing as they see Onnophris' happiness in 
giving his throne to his son Horus before Re in heaven.''}

\item[$g(1,7)$]
$\equiv \Theta_{\mathrm{Algol}}=32\aste$ (Figure \ref{figone}: \SW{Horus})
{\it
``A day of feast of Heaven and of Earth, so too of all people. 
A feast of entering into heaven. The two banks of Horus rejoice.''}

\item[$g(27,3)$]
$\equiv \Theta_{\mathrm{Algol}}=38\aste$ (Figure \ref{figone}: \SW{Horus})
{\it ``Judgement between Horus and Seth. 
Stopping the fight, hunting the rowers, pacifying the raging one. 
Satisfying of the two lords, causing peace to the land. 
The whole of Egypt is given to Horus 
and the whole of desert is given to Seth. 
Coming forth of Thoth who speaks the decree in front of Re.''}

\item[$g(15,11)$]
$\equiv \Theta_{\mathrm{Algol}}=38\aste$ (Figure \ref{figone}: \SW{Horus}) 
{\it
``If you see a thing, it is good. 
Horus hears your words in front of every god and every goddess on this day, 
concerning every good thing you see in your house.''}

\item[$g(1,9)$]
$\equiv \Theta_{\mathrm{Algol}}=51\aste$ (Figure \ref{figone}: \SW{Horus}) 
{\it ``Feast of Horus, the son of Isis. His followers in ...'' }

\item[$g(3,2)$]
$\equiv \Theta_{\mathrm{Algol}}=57\aste$ (Figure \ref{figone}: \SW{Wedjat})   
{\it ``The coming forth of Thoth in the presence of Re in the hidden shrine. He gave the inscription of the pacification of the Wedjat-eye; Hu , Sia and the followers of Maat obey Neith, the crown-goddess in his retinue. If you see anything, 
it will be good on this day.'' } 

\item[$g(7,9)$]
$\equiv \Theta_{\mathrm{Algol}}=88\aste$ (Figure \ref{figone}: \SW{Horus})
{\it ``The crew and followers of Horus have assembled in the foreign land,
to make known that Horus smites him who rebels against his lord. 
Every land is content, their hearts in great joy.''}

\item[$g(28,3)$]
$\equiv \Theta_{\mathrm{Algol}}=164\aste$ (Figure \ref{figone}: \SW{Horus})
{\it 
``The gods are in jubilation and in joy over 
the making of will for Horus, son of Osiris, 
to pacify Onnophris in the underworld. 
Then the land is in feast and the hearts of the gods are pleased. 
If you see anything, it will be good on this day.''}

\item[$g(1,10)$]
$\equiv \Theta_{\mathrm{Algol}}=240\aste$
(Figure \ref{figone}: \SW{Horus}) 
{\it ``Horus... Osiris... Chentechtai... Land''}

\item[$s(26,1)$]
$\equiv \Theta_{\mathrm{Algol}}=253\aste$ (Figure \ref{figtwo}: \SW{Horus})
{\it  \Puna
``Do not do anything on this day. 
This is the day of fighting between Horus and Seth. 
Every man grasped his fellow and they were on their sides as two men. 
They were turned into two ebonies in the netherworld of the lords of Babylon. 
Three days and nights were spent in this manner. 
Then Isis let their harpoons fall. It fell in front of her son Horus. 
Then he called with a loud voice saying he is her son Horus. 
Then Isis called to this harpoon: Loosen, loosen from son Horus! 
Then this harpoon loosened from her son Horus. 
Then she let fall another harpoon in front of her brother Seth. 
He shouted saying he is her brother Seth. 
Then she called to this harpoon: Be strong! Be strong! 
Then this Seth shouted to her many times saying: 
Do I love the stranger more than the brother of the mother? 
Then her heart was greatly saddened and she called to this harpoon:
 Loosen, loosen! Behold the brother of the mother. 
So was this harpoon driven from him. 
They stood up as two men and each turned his back against another. 
Then the majesty of Horus 
was angered against his mother Isis like a panther. 
She placed it in front of him.''}

\item[$s(11,11)$]
$\equiv \Theta_{\mathrm{Algol}}=253\aste$ (Figure \ref{figtwo}: \SW{Horus})
{\it  \Puna
``Re's bringing of the great ones to the booth 
to see what he had seen through the eye of Horus the Elder. 
Then their faces were turned down seeing the eye of Horus 
raging in front of Re. 
Do not perform any ritual in any house on this day.''}

\item[$s(10,6)$]
$\equiv \Theta_{\mathrm{Algol}}=259\aste$ (Figure \ref{figtwo}: \SW{Wedjat})  
{\it  \Puna
``The coming forth of Wedjat into the presence of the favoured ones in Heliopolis. The promotion of the Majesty of the shrines through Mnevis-bull, the herald of Re, whom Maat raised up to Atum.''}

\item[$s(27,8)$]
$\equiv \Theta_{\mathrm{Algol}}=265\aste$ (Figure \ref{figtwo}: \SW{Sakhmet})
{\it  \Puna
``Do not go out of your house until the setting of the sun 
because the majesty of Sakhmet 
violates in ``Tjemehu''\Egy,
where she is walking without anyone nearby.''}

\item[ $g(16,4)$]
$\equiv \Theta_{\mathrm{Algol}}=278\aste$ (Figure \ref{figone}:  \SW{Sakhmet})  
{\it ``This is the day of the feast of Sakhmet and Bastet in Asheru for Re, given in front of Re.''}

\item[$s(13,6)$]
$\equiv \Theta_{\mathrm{Algol}}=278\aste$ (Figure \ref{figtwo}: \SW{Sakhmet})
{\it  \Puna
``Do not go out of your house to any road on this day. 
This is the day of the arrival of Sakhmet of Reheset. 
Their great ``slaughterer-demons'' were let loose 
from Letopolis on this day.''}

\item[$s(7,10)$] 
$\equiv \Theta_{\mathrm{Algol}}=278\aste$ (Figure \ref{figtwo}: \SW{Sakhmet})
{\it  \Puna
``Do not go out of your house to spend time 
until the setting of the sun to the horizon. 
This is the day of the hidden-named ``slaughterer-demons'' of Sakhmet...''}

\item[$g(23,7)$]
$\equiv \Theta_{\mathrm{Algol}}=291\aste$ (Figure \ref{figone}: \SW{Horus}) 
{\it 
``Feast of Horus in Athribis on this day of his years, 
in his great and beautiful images.''}

\item[$g(29,3)$]
$\equiv \Theta_{\mathrm{Algol}}=291\aste$ (Figure \ref{figone}: \SW{Horus}) 
{\it
``Coming forth of the three ancestors inside the Tanenet in front of Ptah, 
beautiful of face, 
while adoring Re of the throne of the truth of the goddess temples. 
The white crown is given to Horus and the red one to Seth. 
Their hearts are pleased upon them.''}

\item[$s(20,9)$]
$\equiv \Theta_{\mathrm{Algol}}=291\aste$ (Figure \ref{figtwo}: \SW{Horus})
{\it  \Puna  ``The judgement of Maat in front of these gods, 
angered on the island of the sanctuary of Letopolis, 
inspected by the majesty of Horus.''}

\item[$g(9,5)$] 
$\equiv \Theta_{\mathrm{Algol}}=303\aste$
(Figure \ref{figone}: \SW{Sakhmet})
{\it ``The gods are joyful over the matter of Sakhmet.
The day of establishing the food offerings and reversion-offering, 
which are pleasing to the gods 
and the {``akh''.''}}

\item[$g(30,10)$]
$\equiv \Theta_{\mathrm{Algol}}=303\aste$ (Figure \ref{figone}: \SW{Wedjat})  
{\it ``The coming forth of Shu to bring back Wedjat, pacified by Thoth on this day. House of Re. House of Osiris. House of Horus.''} 

\item[$g(29,5)$] 
$\equiv \Theta_{\mathrm{Algol}}=309\aste$
(Figure \ref{figone}: \SW{Sakhmet})
{\it ``Appearance in glory in the sight of Hu by Thoth to send this decree southwards
to instruct the two lands of Bastet together with the sole mistress Sakhmet the great. The gods are pleased. If you see anything, it will be good on this day.''}

\item[$g(18,1)$]
$\equiv \Theta_{\mathrm{Algol}}=322\aste$ (Figure \ref{figone}: \SW{Horus})
{\it ``If you see anything, it will be good on this day. 
This is the day of magnifying the majesty of Horus over his brother, 
which they did at the gate.''}

\item[$g(6,9)$]
$\equiv \Theta_{\mathrm{Algol}}=322\aste$ (Figure \ref{figone}: \SW{Wedjat}) 
{\it ``The coming of the great ones from the House of 
Re. Joy on this day, as they catch Wedjat together with their followers. 
If you see anything, 
it will be good on this day.''}  

\end{itemize}

\section{List 2}

\begin{itemize}

\item[$g(1,9)$]
$\equiv \Theta_{\mathrm{Moon}}=0\aste$ (Figure \ref{figthree}: \SW{Horus}) 
{\it ``Feast of Horus, the son of Isis. His followers in ...'' }

\item[$g(1,10)$]
$\equiv \Theta_{\mathrm{Moon}}=5\aste$
(Figure \ref{figthree}: \SW{Horus} and \SW{Osiris}) 
{\it ``Horus... Osiris... Chentechtai... Land''}

\item[$s(5,8)$]
 $\equiv \Theta_{\mathrm{Moon}}=44\aste$ (Figure \ref{figfour}: \SW{Horus})
{\it \Puna ``The majesty of Horus is proceeding while Deshret 
sees his image. 
Every approach to him is met with rage.''}

\item[$g(6,7)$] 
$\equiv \Theta_{\mathrm{Moon}}=51\aste$
(Figure \ref{figthree}: \SW{Osiris})
{\it ``Joy of Osiris at the tomb of Busiris. The coming forth of Anubis, adoration in his wake, likewise he has received all people. (his) adorers (or, adoration) following
him; he has received everybody in the hall. Perform a ritual!''}

\item[$g(9,4)$] 
$\equiv \Theta_{\mathrm{Moon}}=73\aste$
(Figure \ref{figthree}: \SW{Seth})
{\it ``It is the day of doing what Thoth did. 
 'The djed-pillars endure', 
says Re in front of the great ones, whereupon these gods together with Thoth let the enemy of Seth kill himself in front of his sanctuary. 
This is what was done
by the ``slaughterer-demons'' of {Qesret on} this day.'' 
}

\item[$g(7,9)$]
$\equiv \Theta_{\mathrm{Moon}}=73\aste$ (Figure \ref{figthree}: \SW{Horus})
{\it ``The crew and followers of Horus have assembled in the foreign land,
to make known that Horus smites him who rebels against his lord. 
Every land is content, their hearts in great joy.''}

\item[$g(11,4)$] 
$\equiv \Theta_{\mathrm{Moon}}=98\aste$
(Figure \ref{figthree}: \SW{Osiris})
{\it ``Feast of Osiris in Abydos in the 
great ``neshmet-bark''\Egy on this day. 
The dead are in jubilation.''}

\item[$s(12,2)$] 
$\equiv \Theta_{\mathrm{Moon}}=100\aste$
(Figure \ref{figfour}: \SW{Seth})
{\it  \Puna
``This is the day of raising his head by the one who rebelled against his lord. His intent is destroyed and the staff of Seth, son of Nut. His head, who did rebel against his lord, is cut off.''}

\item[$s(13,3)$] 
$\equiv \Theta_{\mathrm{Moon}}=117\aste$
(Figure \ref{figfour}: \SW{Seth} and \SW{Osiris})
{\it  \Puna
``This is the day of severing... the ferryman upon the uncrossable river of snakes... every hall to 
this ``neshmet-bark'' of Osiris, 
sailing southwards to Abydos, to the great city of Onnophris. For he has made his form into one old and small in the 
arms of (translation unknown)... 
given gold as reward to Nemti for fare, saying 'Ferry us to the west!' Then he received it... upon a limb of the divine body, whereupon was this association behind him as an 
army of ``reptiles''\Egy. 
Then they knew Seth had made these gods enter to purify the limb of the divine body. Then they revived it... he came... the enemy behind him on the water. Then they changed their forms into little, small cattle. 
Then these 
gods made a ``pile of corpses''\Egy
and split them entirely. 
Then was taken action upon the tongue of the enemy of Nemti. 
Do not approach the gold in the house of Nemti as far as this day. 
So began the removal of the little, small cattle from the west, 
so began the creation of the herds of little, small cattle as far as this day.  
''}

\item[$g(14,2)$]
$\equiv \Theta_{\mathrm{Moon}}=124\aste$ (Figure \ref{figthree}: \SW{Horus})
{\it
``The day of the majesty of Horus receiving the white crown. 
His Ennead is in a great celebration. 
Make offering to the gods of your city. 
Pacify your akh.''
} 

\item[$s(14,3)$] 
$\equiv \Theta_{\mathrm{Moon}}=129\aste$
(Figure \ref{figfour}: \SW{Osiris})
{\it  \Puna
``Do not do anything on this day. The hearts
of the gods are sad over the action against Osiris' place of embalming and the action of the enemy of Nemti. All born on this day will die of cuts.''}

\item[$s(11,11)$]
$\equiv \Theta_{\mathrm{Moon}}=132\aste$ (Figure \ref{figfour}: \SW{Horus})
{\it  \Puna
``Re's bringing of the great ones to the booth 
to see what he had seen through the eye of Horus the Elder. 
Then their faces were turned down seeing the eye of Horus 
raging in front of Re. 
Do not perform any ritual in any house on this day.''}

\item[$s(11,12)$]
$\equiv \Theta_{\mathrm{Moon}}=137\aste$
(Figure \ref{figfour}: \SW{Seth}) 
{\it  \Puna
``The causers of tumult are in front of the followers of Re, who repelled the confederacy of Seth to the eastern desert.''}

\item[$s(14,5)$] 
$\equiv \Theta_{\mathrm{Moon}}=139\aste$
(Figure \ref{figfour}: \SW{Osiris})
{\it  \Puna
``Weeping of Isis and Nephthys. It is the day of their mourning Osiris in Busiris in remembrance of that which he had seen. Do not listen to singing or music
on this day.''}

\item[$g(16,2)$] 
$\equiv \Theta_{\mathrm{Moon}}=149\aste$
(Figure \ref{figthree}: \SW{Osiris})
{\it ``Feast of Osiris-Onnophris. The gods who are in his attendance are in great celebration. The Ennead is before Re, joyful. If you see anything on this day, it will be good.
''}

\item[$g(13,12)$] 
$\equiv \Theta_{\mathrm{Moon}}=161\aste$
(Figure \ref{figthree}: \SW{Seth} and \SW{Osiris})
{\it ``A holiday because of protecting the son of Osiris. . . 
at the back of the portal by Seth.''}

\item[$g(18,1)$]
$\equiv \Theta_{\mathrm{Moon}}=168\aste$ (Figure \ref{figthree}: \SW{Horus})
{\it ``If you see anything, it will be good on this day. 
This is the day of magnifying the majesty of Horus over his brother, 
which they did at the gate.''} 

\item[$s(18,3)$] 
$\equiv \Theta_{\mathrm{Moon}}=178\aste$
(Figure \ref{figfour}: \SW{Seth})
{\it  \Puna
''This is the day of tumult by the children of Geb: Seth and his sister Nephthys. Do not approach any road until the deed is done on this day.''}

\item[$g(15,11)$]
$\equiv \Theta_{\mathrm{Moon}}=180\aste$ (Figure \ref{figthree}: \SW{Horus}) 
{\it
``If you see a thing, it is good. 
Horus hears your words in front of every god and every goddess on this day, 
concerning every good thing you see in your house.''}

\item[$g(17,6)$] 
$\equiv \Theta_{\mathrm{Moon}}=180\aste$
(Figure \ref{figthree}: \SW{Osiris})
{\it ``This is the day of bringing to the embalming place of Osiris those offerings which have been placed in the hands of Anubis.''}

\item[$s(17,7)$] 
$\equiv \Theta_{\mathrm{Moon}}=185\aste$
(Figure \ref{figfour}: \SW{Seth})
{\it  \Puna
``Do not speak the name of Seth on this day. Who in his lack of knowledge pronounces his name, he will not stop fighting in his house of eternity.''}

\item[$s(17,8)$] 
$\equiv \Theta_{\mathrm{Moon}}=190\aste$
(Figure \ref{figfour}: \SW{Seth})
{\it  \Puna
``The going of Seth, son of Nut, to the brawlers that have been reckoned on his day. These gods became aware of him, they repelled his followers and none of them remained.''}

\item[$s(19,4)$] 
$\equiv \Theta_{\mathrm{Moon}}=195\aste$
(Figure \ref{figfour}: \SW{Osiris})
{\it  \Puna
``Stealing of property 
inside the Deshret ``temple building''\Egy. 
Making of ointment for Osiris in front of the funerary workshop. 
Do not taste bread or beer on this day. Drink the juice of grapes
until Re sets.''}

\item[$s(20,2)$]  
$\equiv \Theta_{\mathrm{Moon}}=197\aste$
(Figure \ref{figfour}: \SW{Seth})
{\it  \Puna
``This is the day of giving food-offerings in front of Re and followers by Thoth. The act in there was done accordingly. The rebels against their lord were overthrown. Then they lifted up Seth, son of Nut; 
so they became lowered by the gods.''}

\item[$s(20,9)$]
$\equiv \Theta_{\mathrm{Moon}}=231\aste$ (Figure \ref{figfour}: \SW{Horus})
{\it  \Puna
``The judgement of Maat in front of these gods, 
angered on the island of the sanctuary of Letopolis, 
inspected by the majesty of Horus.''}

\item[$g(19,12)$]
$\equiv \Theta_{\mathrm{Moon}}=234\aste$ (Figure \ref{figthree}: \SW{Horus})
{\it 
``Feast for your god! 
Propitiate your ``akh'', 
because this eye of Horus has come, is complete, 
is uninjured and there is no claim against it.''}

\item[$g(24,3)$]
$\equiv \Theta_{\mathrm{Moon}}=251\aste$ (Figure \ref{figthree}: \SW{Horus})
{\it
``The arrival of Isis joyful and Nephthys 
rejoicing as they see Onnophris' happiness in 
giving his throne to his son Horus before Re in heaven.''}

\item[$g(23,7)$]
$\equiv \Theta_{\mathrm{Moon}}=258\aste$ (Figure \ref{figthree}: \SW{Horus}) 
{\it 
``Feast of Horus in Athribis on this day of his years, 
in his great and beautiful images.''}

\item[$s(26,1)$] 
$\equiv \Theta_{\mathrm{Moon}}=266\aste$
(Figure \ref{figfour}: \SW{Horus} and \SW{Seth})
{\it  \Puna
``Do not do anything on this day. 
This is the day of fighting between Horus and Seth. 
Every man grasped his fellow and they were on their sides as two men. 
They were turned into two ebonies in the netherworld of the lords of Babylon. 
Three days and nights were spent in this manner. 
Then Isis let their harpoons fall. It fell in front of her son Horus. 
Then he called with a loud voice saying he is her son Horus. 
Then Isis called to this harpoon: Loosen, loosen from son Horus! 
Then this harpoon loosened from her son Horus. 
Then she let fall another harpoon in front of her brother Seth. 
He shouted saying he is her brother Seth. 
Then she called to this harpoon: Be strong! Be strong! 
Then this Seth shouted to her many times saying: 
Do I love the stranger more than the brother of the mother? 
Then her heart was greatly saddened and she called to this harpoon:
 Loosen, loosen! Behold the brother of the mother. 
So was this harpoon driven from him. 
They stood up as two men and each turned his back against another. 
Then the majesty of Horus 
was angered against his mother Isis like a panther. 
She placed it in front of him.''}

\item[$s(24,8)$] 
$\equiv \Theta_{\mathrm{Moon}}=275\aste$
(Figure \ref{figfour}: \SW{Seth})
{\it  \Puna
``Do not pronounce the name of Seth. Do not raise your voice on this day. This is the day of Onnophris. As to anyone who pronounces his name in ignorance, he shall not cease fighting in his house for ever.''}

\item[$g(27,1)$] 
$\equiv \Theta_{\mathrm{Moon}}=278\aste$
(Figure \ref{figthree}: \SW{Horus} and \SW{Seth})
{\it ``Peace between Horus and Seth. Do not kill any snakes on this day. 
Make a good day!''}

\item[$g(27,3)$] 
$\equiv \Theta_{\mathrm{Moon}}=287\aste$
(Figure \ref{figthree}: \SW{Horus} and \SW{Seth})
{\it 
``Judgement between Horus and Seth. 
Stopping the fight, hunting the rowers, pacifying the raging one. 
Satisfying of the two lords, causing peace to the land. 
The whole of Egypt is given to Horus 
and the whole of desert is given to Seth. 
Coming forth of Thoth who speaks the decree in front of Re.''
}

\item[$g(28,3)$] 
$\equiv \Theta_{\mathrm{Moon}}=300\aste$
(Figure \ref{figthree}: \SW{Horus} and \SW{Osiris})
{\it 
``The gods are in jubilation and in joy over 
the making of will for Horus, son of Osiris, 
to pacify Onnophris in the underworld. 
Then the land is in feast and the hearts of the gods are pleased. 
If you see anything, it will be good on this day.''}

\item[$g(29,3)$] 
$\equiv \Theta_{\mathrm{Moon}}=312\aste$
(Figure \ref{figthree}: \SW{Horus} and \SW{Seth})
{\it ``Coming forth of the three ancestors inside the Tanenet in front of Ptah, beautiful of face, while adoring Re of the throne of truth of 
the goddess temples. The white crown is given to Horus and the red one to Seth. Their hearts are pleased upon them.''}

\item[$g(28,7)$]
$\equiv \Theta_{\mathrm{Moon}}=319\aste$ 
(Figure \ref{figthree}: \SW{Osiris})
{\it ``Feast of Osiris in Abydos. The majesty of Onnophris raised up
the willow.''}

\item[$g(1,7)$]
$\equiv \Theta_{\mathrm{Moon}}=351\aste$ (Figure \ref{figthree}: \SW{Horus})
{\it
``A day of feast of Heaven and of Earth, so too of all people. 
A feast of entering into heaven. The two banks of Horus rejoice.''}
\end{itemize}

\end{document}